\newcommand\beq{\begin{eqnarray}}
\newcommand\eeq{\end{eqnarray}}
\newcommand\missingET{E_T^{\rm miss}}
\newcommand\missET{E_T^{\rm miss}}

\def\lsim{\mathrel{\rlap{\lower4pt\hbox{$\sim$}}
    \raise1pt\hbox{$<$}}}                
\def\gsim{\mathrel{\rlap{\lower4pt\hbox{$\sim$}}
    \raise1pt\hbox{$>$}}}

\documentclass[
amsmath,
prd,nofootinbib,floatfix,11pt,
]{revtex4}

\allowdisplaybreaks
\usepackage{graphicx}
\usepackage{setspace}

\begin{document}
\renewcommand{\theequation}{\arabic{section}.\arabic{equation}}
\begin{flushright}
NSF-KITP-08-107
\end{flushright}

\title{\Large%
\baselineskip=21pt
Exploring compressed supersymmetry with same-sign top quarks 
at the Large Hadron Collider}

\author{Stephen P. Martin}
\affiliation{
{\it Department of Physics, Northern Illinois University, DeKalb IL 60115,} 
\\
{\it Fermi National Accelerator Laboratory, P.O. Box 500, Batavia IL 60510,}
and\\
{\it Kavli Institute for Theoretical Physics, University of California, 
Santa Barbara CA 93106-4030
}
}

\begin{abstract}\normalsize \baselineskip=15pt 
In compressed supersymmetry, a light top squark naturally mediates 
efficient neutralino pair annihilation to govern the thermal relic 
abundance of dark matter. I study the LHC signal of same-sign leptonic 
top-quark decays from gluino and squark production, which follows from 
gluino decays to top plus stop followed by the stop decaying to a charm 
quark and the LSP in these models. Measurements of the numbers of jets 
with heavy-flavor tags in the same-sign lepton events can be used to 
confirm the origin of the signal. 
Summed transverse momentum observables provide an 
estimate of an effective superpartner mass, which is correlated with the 
gluino mass. Measurements of invariant mass endpoints from the visible 
products of gluino decays do not allow direct determination of 
superpartner masses, but can place constraints on them, including lower 
bounds on the gluino mass as a function of the top-squark mass.
\end{abstract}


\maketitle

\tableofcontents

\vfill\eject
\baselineskip=14.9pt

\setcounter{footnote}{1}
\setcounter{page}{2}
\setcounter{figure}{0}
\setcounter{table}{0}

\section{Introduction}
\label{sec:intro}
\setcounter{equation}{0}
\setcounter{footnote}{1}

If supersymmetry \cite{SUSYreviews} is the solution to the hierarchy
problem associated with the small ratio of the electroweak scale to the
Planck scale, then some of the superpartners should be discovered at the
impending CERN Large Hadron Collider (LHC). While the essential idea of
supersymmetry as a symmetry connecting fermion and boson degrees of
freedom is quite predictive, the unknown features of the supersymmetry
breaking mechanism allow for a diverse variety of possibilities for the
LHC signals of superpartner production and decay \cite{ATLASTDR,CMSTDR}. 

The purpose of this paper is to study some of the distinctive LHC signals
particular to the ``compressed supersymmetry" scenario proposed in
ref.~\cite{compressedSUSY}, which is motivated both by the supersymmetric
little hierarchy problem and by the cold dark matter relic abundance
obtained by WMAP, SDSS and other experiments \cite{WMAP,SDSS}.  This model
scenario follows from assuming that the ratio of the running gluino and
wino mass parameters, $M_3/M_2$, is smaller than 1 near the GUT scale,
unlike the assumption of the well-studied ``minimal supergravity" 
(mSUGRA) framework.  (Models with non-unified gaugino masses have recently
attracted renewed interest, see for example
\cite{Ellis:1985jn}-\cite{Everett:2008ey}, due to their ability to
incorporate novel LHC phenomenology and dark-matter physics.) As a result,
the ratio of the physical masses of the heaviest and the lightest
superpartners is much less than in mSUGRA, because the gluino mass feeds
into the other superpartner masses by renormalization group evolution.
Another characteristic feature is that the pair annihilation of the
lightest supersymmetric particles (LSPs) in the early universe can
naturally proceed dominantly through $\tilde N_1 \tilde N_1 \rightarrow t
\overline t$, mediated by top-squark exchange. This is due to the fact
that the stop-LSP mass difference is naturally not too large, particularly
in models that have enough top-squark mixing to evade the Higgs scalar
boson mass bound from LEP. 

In compressed supersymmetry, the superpartner masses are typically all
less than 1 TeV in the available parameter space. This means that the
initial evidence for supersymmetry should follow quickly from the classic
jets with missing transverse energy $(\missET)$ signal, as soon as the
systematic difficulties associated with understanding missing energy as
manifested in the LHC detectors are conquered.  One can then turn
attention to those features of the signal that might distinguish it from
the usual mSUGRA models. In much of the parameter space in compressed
supersymmetry that predicts the observed thermal relic abundance of dark
matter, the mass difference between the top squark and the neutralino LSP
$\tilde N_1$ is less than 85 GeV, so that the flavor-preserving decays
including
$\tilde t_1 \rightarrow t \tilde N_1$ and 
$\tilde t_1 \rightarrow b \tilde C_1$ and 
$\tilde t_1 \rightarrow b W \tilde N_1$ 
are kinematically forbidden. In this paper, I will assume that
the flavor-violating 2-body decay
\beq
\tilde t_1 \rightarrow c \tilde N_1
\label{eq:twobody}
\eeq
dominates
over the remaining possibility \cite{Hikasa:1987db,Boehm:1999tr}, 
the 4-body decay 
$\tilde t_1 \rightarrow b f \bar f' \tilde N_1$.
(For more on this assumption, see the next section.)
Assuming the top squark is lighter than the gluino, one has: 
\beq
\tilde g \rightarrow  
\Biggl 
\{ \begin{array}{ll}
t \tilde t_1^*
\qquad (50\%)
\\
\bar t \tilde t_1
\qquad (50\%),
\end{array}
\Biggr.
\eeq 
due to the Majorana nature of the gluino,
leading to
\beq
pp \rightarrow \tilde g \tilde g \rightarrow
\left \{ \begin{array}{lll}
t \,t\, \overline c \,\overline c \tilde N_1 \tilde N_1\qquad(25\%)
\\
\overline t\, \overline t\, c \, c  \tilde N_1 \tilde N_1\qquad(25\%)
\\
t \,\overline t\, c\, \overline c \tilde N_1 \tilde N_1\qquad (50\%).
\end{array}
\right.
\label{eq:likesigntops}
\eeq
When both of the same-sign top quarks of the first two cases decay
leptonically, one obtains a distinctive detector signal of two same-sign
leptons, two potentially $b$-tagged jets, two or more additional jets, and
missing energy from the LSPs and neutrinos: 
\beq
pp \rightarrow \ell^\pm\ell^{\prime \pm} bbjj+\missET .
\eeq
This is a special case of the well-known same-sign dilepton signature for
Majorana gluino (or gaugino) production in supersymmetry
\cite{SSdileptons}. This LHC signal for gluino pairs was proposed and
studied in some detail, in the context of models with much lighter top
squarks ($m_{\tilde t_1} < m_t$), in ref.~\cite{Kraml:2005kb}. Adding to
this signal in compressed supersymmetry will be events in which squarks
are produced, giving extra jets in the final state when they decay to the
gluino.  The presence of same-sign leptons provides for a strongly
suppressed Standard Model background compared to other missing energy
signals, and this is further aided by requiring two $b$-tagged jets. 

In this paper, I will consider the properties of the LHC events that
conform to this signal in compressed supersymmetry.  (Other studies
of the LHC phenomenology of compressed supersymmetry are found in
\cite{Baer:2007uz}-\cite{Martin:2008sv}.) Section \ref{sec:modelline}
defines a model line for study, a one-parameter slice of model space with
the free parameter corresponding to the gaugino mass scale. I also discuss
some of the prominent properties of the superpartner mass spectrum of this
model line that make it qualitatively different from mSUGRA models.
Section \ref{sec:sstops} describes an event selection for the
$\ell^\pm\ell^{\prime \pm} bbjj+\missET$, and the features of the
resulting signal events. Section \ref{sec:ptsums} considers
mass-estimating observables based on the scalar sum of transverse momentum
of detector objects, while section \ref{sec:endpoints} studies kinematic
endpoints of the invariant masses of visible products of the gluino decay.
Section \ref{sec:outlook} contains some concluding remarks. 

\section{A compressed supersymmetry model line}
\label{sec:modelline}
\setcounter{equation}{0}
\setcounter{footnote}{1}

One simple realization of compressed supersymmetry is obtained by
supposing that the running bino, wino, and gluino masses are parameterized
at $M_{\rm GUT}$ by: 
\beq
M_1 &=& m_{1/2} (1 + C_{24}),
\label{eq:binoM}
\\
M_2 &=& m_{1/2} (1 + 3 C_{24}),
\\
M_3 &=& m_{1/2} (1 - 2 C_{24}),
\label{eq:gluinoM}
\eeq
corresponding to an $F$-term source for supersymmetry breaking in a linear
combination of the singlet and adjoint representations of $SU(5)$
\cite{Ellis:1985jn}-\cite{Anderson:1999ui}.  Merely for simplicity, I also
assume a common scalar mass $m_0$ and scalar trilinear coupling $A_0$,
both at $M_{\rm GUT}$. The other parameters defining the model are
$\tan\beta$ and the phase of the $\mu$ parameter, which is taken to be
real. I use {\tt SOFTSUSY 2.0.11} \cite{softsusy} to generate the
superpartner spectrum. To define the model line for study here, let: 
\beq
C_{24} = 0.21,\qquad A_0/M_1 = -1,\qquad \tan\beta = 10,\qquad \mu > 0,
\eeq
with $M_1$ (or equivalently $m_{1/2}$) taken as the single varying
parameter of the model line.  (Here and in the following, $M_1$ is used to
denote the running bino mass parameter at $M_{\rm GUT}$, not at the
electroweak scale.) For each value of $M_1$, the parameter $m_0$ is
obtained by imposing as a requirement that the predicted dark matter relic
abundance (obtained using the program {\tt micrOMEGAs 2.0.1}
\cite{micrOMEGAs}) satisfies $\Omega_{\rm DM} h^2 = 0.11$
\cite{WMAP,SDSS}. The resulting values of $m_0$ are not too large, ranging
from 210 to 380 GeV (and always less than the wino and bino masses at the
GUT scale) for the model line when the physical gluino mass is less than 1
TeV. 

With strictly flavor-conserving boundary conditions for the soft 
supersymmetry-breaking interactions at the GUT scale, the 2-body decay 
$\tilde t_1 \rightarrow c \tilde N_1$ and the 4-body decay $\tilde t_1 
\rightarrow b f \bar f' \tilde N_1$ would have roughly comparable partial 
decay widths on this model line. Using SDECAY \cite{SDECAY}, one finds 
that BR$(\tilde t_1 \rightarrow b f \bar f' \tilde N_1)$ would range from 
a few percent (for a small mass difference $m_{\tilde t_1} - m_{\tilde 
N_1} \approx 30$ GeV), to nearly 90\% (for the largest mass difference of 
about 70 GeV).\footnote{The most important contribution to the 4-body 
decay partial width in this model line comes from the Feynman graph with 
virtual $W$ and top-quark exchange, unlike in most other models 
considered in the literature (see, for example, 
\cite{Boehm:1999tr,Das:2001kd,Hiller:2008wp}), where diagrams with $W$ 
and chargino or slepton and chargino exchange dominate.} However, the 
strict minimal flavor violation assumption on which this is based is 
notoriously unmotivated by theory, except in models with special features 
like gauge-mediated supersymmetry breaking. A small amount of non-minimal 
flavor violation results in the 2-body decay $\tilde t_1 \rightarrow c 
\tilde N_1$ dominating, as assumed here. Writing the effective 
stop-charm-neutralino interaction at the weak scale as ${\cal L} = 
-\tilde t_1^* (y_L c_L \tilde N_1 + y_R c_R \tilde N_1) $, I find that 
even for the worst-case point on the model line, BR$(\tilde t_1 
\rightarrow c \tilde N_1) > 95\%$ provided that $({y_L^2 + y_R^2})^{1/2} 
> 8\times 10^{-4}$. This would follow, for example, from a small 
off-diagonal right-handed up squark squared mass parameter at the GUT 
scale $m^2_{\tilde c_R^* \tilde t_R}/m_0^2 > 0.007$, with no danger of 
conflict with present flavor experiments.

\begin{figure}[!tp]
\begin{minipage}[]{0.37\linewidth}
\caption{\label{fig:spectrum500}
The mass spectrum for a sample point on the model line described in the 
text, with $M_1 = 500$ GeV at the GUT scale, $C_{24} = 0.21$,
$A_0/M_1 = -1$, $m_0 = 314$ GeV, $\mu = 361$ GeV, and $\tan\beta = 10$.
The columns contain, from left to right,
Higgs scalar bosons, neutralinos,
charginos, the gluino, first and 
second family squarks and sleptons, and third
family squarks and sleptons.}
\end{minipage}
\begin{minipage}[]{0.62\linewidth}
\begin{flushright}
\includegraphics[width=9.4cm,angle=0]{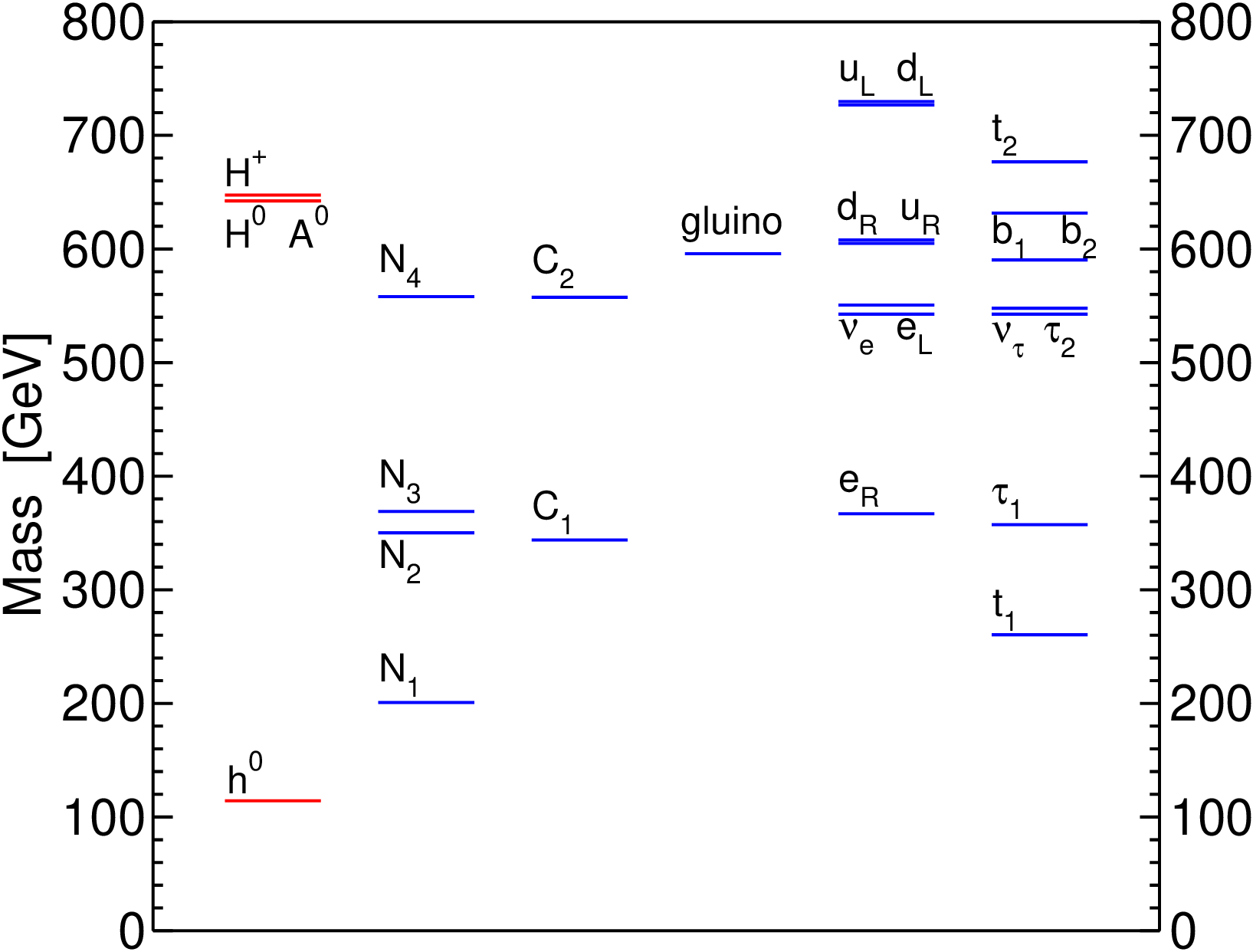}
\end{flushright}
\end{minipage}
\end{figure}

The superpartner and Higgs boson mass spectrum for a representative point 
on the model line with $M_1 = 500$ GeV is shown in figure 
\ref{fig:spectrum500}.
The ratio of masses of the heaviest and lightest superpartners in this
model is 3.6, almost a factor of 2 smaller than is obtainable in mSUGRA
models even with small $m_0$.The neutralino LSP is heavier than the top
quark, allowing $\tilde N_1 \tilde N_1 \rightarrow t \overline t$. The top
squark is the next-to-lightest superpartner. Another distinctive feature
is that $\tilde C_1$ and $\tilde N_{2,3}$ are higgsino-like states, due to
the fact that $\mu$ is only 361 GeV, again much smaller than found in
mSUGRA models for comparable gluino and squark masses.  (The relatively
small value of $|\mu|$ is a sign of the reduced fine-tuning found in
models with a small ratio $M_3/M_2$ at the GUT scale as pointed out long
ago in \cite{KaneKing}.) The heaviest neutralino and chargino states
$\tilde N_4$ and $\tilde C_2$ are wino-like. The sleptons turn out to be
too heavy to play a significant role in LHC physics.

The gluino always decays to top and stop in this model, and the stop
always decays to a charm quark and LSP. 
Other important decay modes are, for left-handed squarks: 
\beq
&&
\tilde u_L
\rightarrow 
u \tilde g\>\> (71\%), 
\quad
d \tilde C_2\>\> (13\%), 
\quad
u \tilde N_4\>\> (6\%),
\quad
d \tilde C_1\>\> (6\%), 
\\
&&
\tilde d_L
\rightarrow
d \tilde g\>\> (73\%),
\quad
u \tilde C_2\>\> (14\%),
\quad
d \tilde N_4\>\> (7\%),
\quad
u \tilde C_1\>\> (3\%),
\eeq
and for right-handed squarks:
\beq
&&
\tilde u_R \rightarrow 
u \tilde N_1\>\> (92\%),
\quad
u \tilde g\>\> (5\%),
\quad
u \tilde N_2\>\> (3\%),
\\
&&
\tilde d_R \rightarrow
d \tilde N_1\>\> (85\%),
\quad
d \tilde g\>\> (12\%),
\quad
d \tilde N_2\>\> (3\%) .
\eeq
Thus left-handed squarks are a plenteous source of gluinos, while
right-handed squarks mostly decay directly to the LSP.  Subdominant decays
produce some neutralinos and charginos, which nearly always decay into
on-shell $W$, $Z$, and $h$ bosons or through top squarks. For the
higgsino-like states: 
\beq
&&
\tilde N_2 \rightarrow 
h \tilde N_1\>\> (90\%),
\quad
Z \tilde N_1\>\> (10\%),
\\
&&
\tilde N_3 \rightarrow 
h \tilde N_1\>\> (97\%),
\quad
Z \tilde N_1\>\> (3\%),
\\
&&
\tilde C_1 \rightarrow 
b \tilde t_1\>\> (91\%),
\quad\,
W \tilde N_1\>\> (9\%),
\eeq
and for the heavier, wino-like, states:
\beq
&&
\tilde N_4 \rightarrow 
W \tilde C_1\>\> (51\%),
\quad
h \tilde N_2\>\> (20\%),
\quad
Z \tilde N_3\>\> (20\%),
\quad
t \tilde t_1 \>\> (8\%),
\\
&&
\tilde C_2 \rightarrow 
W \tilde N_2\>\> (30\%),
\quad
W \tilde N_3\>\> (21\%),
\quad
Z \tilde C_1\>\> (25\%),
\quad
h \tilde C_1\>\> (21\%).
\eeq
An important consequence of these decays is that one cannot find dilepton
mass edges of the type used in \cite{dileptons}-\cite{Lester:2005je} to
obtain information about the superpartner mass spectrum. The only isolated
leptons come from on-shell $W$ and $Z$ decays, since two-body spoiler decays 
are always allowed. Furthermore, sleptons completely decouple from the cascade
decays, because they are too heavy. These features are qualitatively
maintained along the entire model line. 

\begin{figure}[!tp]
\begin{minipage}[]{0.37\linewidth}
\caption{\label{fig:massdiff}
The mass difference $m_{\tilde t_1} - m_{\tilde N_1}$ as a function
of $m_{\tilde g}$, for the model line described in the text. The solid line
is the model line with $\Omega_{\rm DM} h^2 = 0.11$, and the shaded region
denotes the approximate region favored by the 
thermal relic abundance constraints. 
The model line is cut off on the left by the LEP2 Higgs mass constraint.}
\end{minipage}
\begin{minipage}[]{0.62\linewidth}
\begin{flushright}
\includegraphics[width=9.4cm,angle=0]{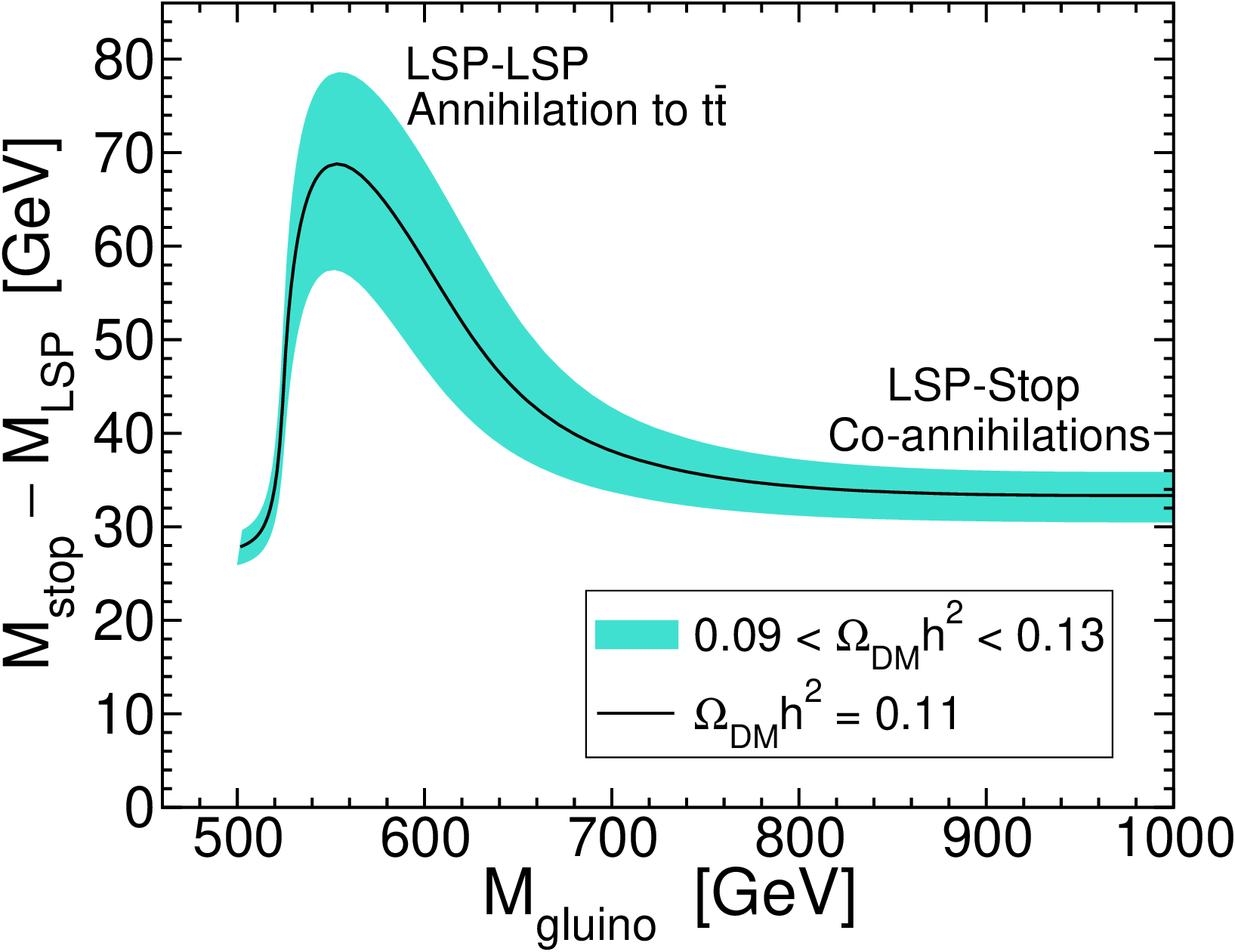}
\end{flushright}
\end{minipage}
\end{figure}

Varying $M_1$, one finds that $M_1 > 417$ GeV at the GUT scale is required
to satisfy the LEP bound on the Higgs mass, taken here to be $m_h > 113$
GeV because of the theoretical uncertainty on the Higgs mass prediction.
As $M_1$ increases, the gluino and LSP masses increase approximately in
direct proportion, while the top-squark mass stays between 30 and 70 GeV
heavier than the neutralino LSP. This is shown in figure
\ref{fig:massdiff}, which plots the mass difference $m_{\tilde t_1} -
m_{\tilde N_1}$ as a function of $m_{\tilde g}$ for the model line. 
The bulge region where the stop-LSP mass difference is relatively large,
with $m_{\tilde g}$ between about 525 and 650 GeV, is characterized by
having $\tilde N_1 \tilde N_1 \rightarrow t \overline t$ due to $\tilde
t_1$ exchange as the dominant annihilation effect in determining the dark
matter thermal relic abundance.  Note that varying $m_0$ to obtain
$\Omega_{\rm DM} h^2$ anywhere within the allowed range $0.11\pm0.02$
would not change the fact that $m_{\tilde t_1} - m_{\tilde N_1} < m_W + m_b$
for this model line. This stop-mediated annihilation region is
continuously connected in parameter space to more fine-tuned models in
which the $\tilde t_1$, $\tilde N_1$ mass difference is just right to
allow efficient stop-neutralino co-annihilations,\footnote{An even more
fine-tuned stop-neutralino co-annihilation region can also be found
\cite{stopcoannihilation} in mSUGRA models.} for $m_{\tilde g}$ less than
about 525 GeV and greater than about 650 GeV. An important consequence of
the larger stop-LSP mass difference in the dark matter
annihilation-to-tops bulge region is that the LHC signal efficiency will
be increased compared to the co-annihilation regions on either side, since
the jets from the decay $\tilde t_1 \rightarrow c \tilde N_1$ tend to have
higher $p_T$. 

\begin{figure}[!tp]
\begin{minipage}[]{0.37\linewidth}
\caption{\label{fig:NLOsigma}
The NLO production cross-section for superpartner pairs
in $pp$ collisions at $\sqrt{s} = 14$ TeV for 
selected points along the model line described in the text, 
as a function of the gluino mass. 
Prospino2 \cite{prospino} was used.
The most important contributions, from stop pair production 
($\tilde t_1 \tilde t_1^*$),
gluino-squark production ($\tilde g \tilde q$ and $\tilde g \tilde q^*$), 
gluino pair production ($\tilde g \tilde g$), and squark pair 
production 
($\tilde q \tilde q$ and $\tilde q \tilde q^*$ and $\tilde q^* \tilde q^*$) 
are also shown separately.}
\end{minipage}
\begin{minipage}[]{0.62\linewidth}
\begin{flushright}
\includegraphics[width=9.4cm,angle=0]{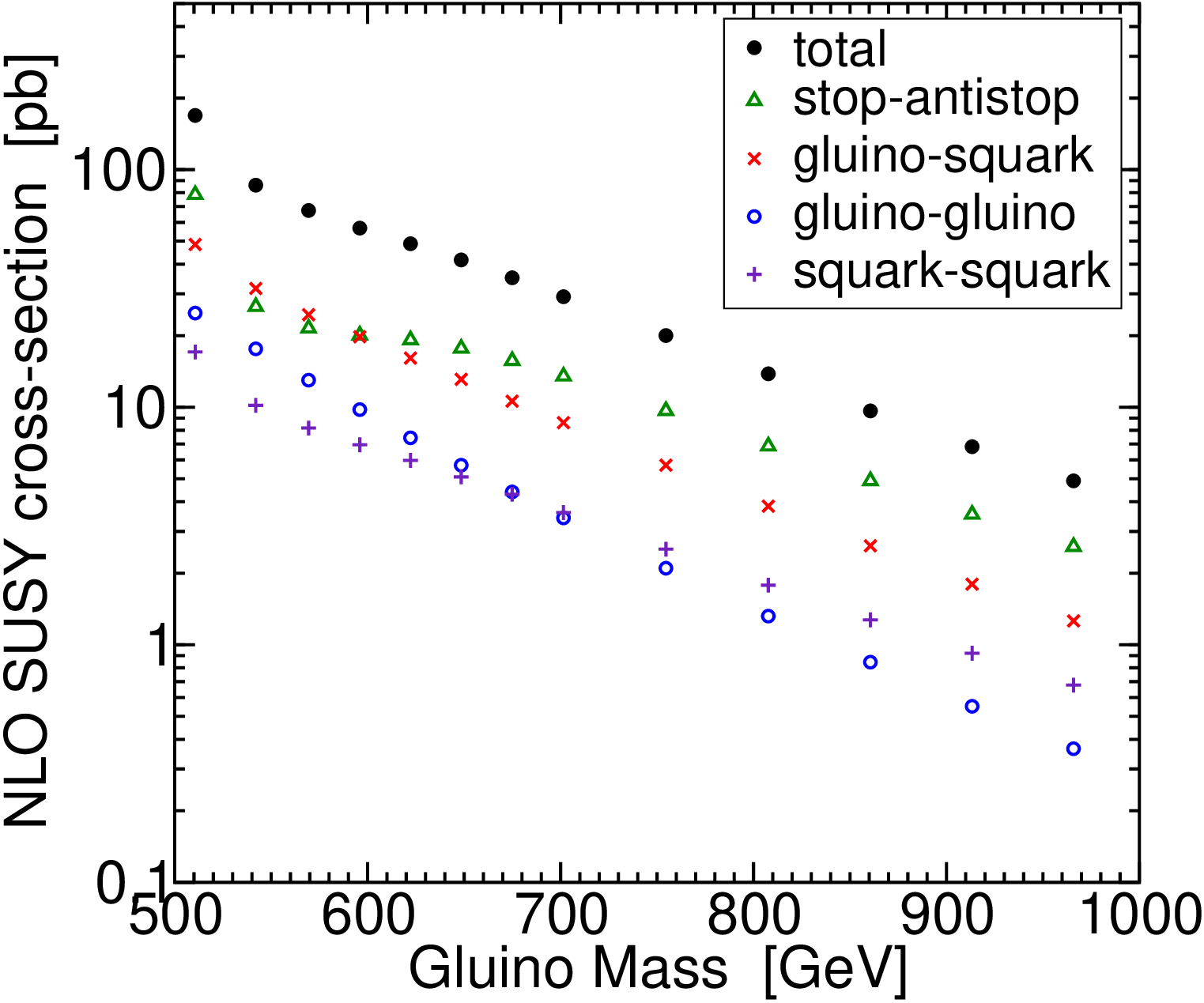}
\end{flushright}
\end{minipage}
\end{figure}

The superpartner production cross-sections are dominated by gluino and
squark production.  The next-to-leading order total cross-sections for
this model line are shown in figure \ref{fig:NLOsigma}, computed using
Prospino2 \cite{prospino}. 
The largest single source of supersymmetric events is $pp\rightarrow
\tilde t_1 \tilde t_1^*$, which is of order 25 pb throughout the bulge
region, and falls rather slowly with the gluino mass along the model line.
However, this leads to the very difficult signal of two often low-$p_T$
charm jets and little missing energy. I have checked that after realistic
cuts to remove QCD and detector backgrounds (see for example
\cite{CMSTDR,Hubisz:2008gg}), the low efficiency of the $\tilde t_1 \tilde
t_1^*$ signal will lead to it being buried beneath the other squark and
gluino sources, so it is not possible to infer the existence of the light
top squark from this direct production process. The total gluino-squark
associated production $pp \rightarrow \tilde g \tilde q$ plus $\tilde g
\tilde q^*$, summed over quark flavors, is of order tens of picobarns
throughout the bulge region. Gluino pair production and (anti-)squark pair
production both contribute of order 10 pb in the bulge region, with the
former falling somewhat more steeply with increasing mass.  The production
and decays of gluinos and squarks in this scenario should easily allow for
early discovery 
lepton+jets+$\missET$ channels (see, for example,
refs.~\cite{ATLASTDR,CMSTDR,Baer:2003wx} for comparable mSUGRA studies) at
the LHC. 

Sleptons decouple from practical LHC physics in many compressed
supersymmetry models, and in particular for the model line studied here.
For example, for the model line point with $M_1 = 500$ GeV shown in figure
\ref{fig:spectrum500}, the total direct production cross-section of
sleptons and sneutrinos before any cuts or efficiencies is only about 6
fb, compared to much larger backgrounds from $WW$ production and other
sources. As noted above, sleptons also extremely efficiently decouple from
decay chains of heavier superpartners.  Charginos and neutralinos (other
than the LSP) do appear, but only in subdominant decay modes of the
squarks. Their direct production rates are quite small compared to the
gluino and squark rates.  For example, for the model shown in figure
\ref{fig:spectrum500}, one obtains a total of (129, 41, 32) fb for,
respectively, ($\tilde C_i^\pm \tilde N_j$, $\tilde C_i^+ \tilde C_j^-$,
$\tilde N_i \tilde N_j$) production. Neutralinos and charginos produced in
association with gluinos and squarks adds another 500 fb. These rates are
quite small compared to the $56$ pb total gluino and squark production
rate, and involve a wide variety of dissimilar final states without strong
distinguishing features.  Furthermore, these do not yield dilepton mass
edges, as noted above. Unfortunately, finding out any information about
the superpartners other than the squarks, gluino, and LSP from direct
observation appears to be a daunting challenge at the LHC in this
scenario. 

There are several ways of gaining information about the gluino and squark
mass spectrum from the early discovery inclusive jets + leptons +
$\missET$ signal, including for example $m_{T2}$ and similar variables
\cite{mTtwoA,mTtwoB} and the multiplicity of $b$-tagged jets. However, the
presence of non-negligible backgrounds that will have to be understood
from LHC data puts these methods beyond the scope of the present paper.
Instead, I will concentrate on tools that use the lower rate but
potentially very low-background signal with same-sign dileptons. 

\section{Signal from same-sign leptonic top decays}
\label{sec:sstops}
\setcounter{equation}{0}
\setcounter{footnote}{1}

To define a signal for same-sign $\ell^\pm\ell^{\prime \pm} bb jj +
\missET$ events, I used MadGraph/MadEvent \cite{MGME} for event
generation, interfaced to Pythia \cite{Pythia} and then PGS4 \cite{PGS}
for detector simulation using CMS-like parameters.  Events are selected by
requiring the following from objects generated by PGS: 
\begin{itemize}
\item exactly two same-sign isolated leptons $(\ell = e,\mu)$ with $p_T > 
20$ GeV and $|\eta| < 2.4$.
\item at least two $b$-tagged jets each with $p_T > 50$ GeV 
(with $|\eta| < 1.75$ required by PGS).
\item at least two more jets with $p_T > 50,35$ GeV 
(with $|\eta| < 3.1$
required by PGS).
\item at least one pairing of each of the 
two leptons with a distinct $b$-tagged jet,
with each pair having invariant mass
consistent with leptonic top decay: $m(b\ell) < 160$ GeV.
\item $\missET > 100$ GeV.
\end{itemize}
(These cuts are very similar to those used in ref.~\cite{Kraml:2005kb}.)
The uncorrected jet momenta from PGS are used.
The reason for the cut on the $b\ell$ invariant mass is that the 
parton-level kinematic endpoint is:
\beq
m(b\ell)_{\rm max} &=& \sqrt{m^2_t - m_W^2},
\eeq
nominally about $153$ GeV using $m_t = 172.7$ GeV. I use the higher value
of 160 GeV for the cut in order to partially take into account the effects
of smearing of the $b$-jet energies. The $b$-tag is actually a heavy
flavor tag, which in PGS has an efficiency for high-$p_T$ central jets of
approximately 50\% for true $b$ jets, 13\% for $c$ jets, and 1\% for
$g,u,d,s$ jets.  Each of the leptons and jets are required to be isolated
from each other by $\Delta R = \sqrt{(\Delta \phi)^2 + (\Delta \eta)^2} >
0.4$. Also, muons that are not isolated from a jet are absorbed into the
jet, if the summed $p_T$ (excluding the muon itself) in a cone of $\Delta
R=0.4$ around the muon exceeds 5 GeV, or if the ratio of the $p_T$ in a
$3\times 3$ grid of calorimeter cells around the muon to the $p_T$ of the
muon itself exceeds 0.1. 

\begin{figure}[!tbp]
\begin{minipage}[]{0.37\linewidth}
\caption{\label{fig:SSsig}
The number of LHC signal events with two 
same-sign leptons, two $b$ tags, and
two additional jets,
per fb$^{-1}$, after the cuts
described in the text, for the model line described in section
\ref{sec:modelline}.}
\end{minipage}
\begin{minipage}[]{0.62\linewidth}
\begin{flushright}
\includegraphics[width=9.2cm,angle=0]{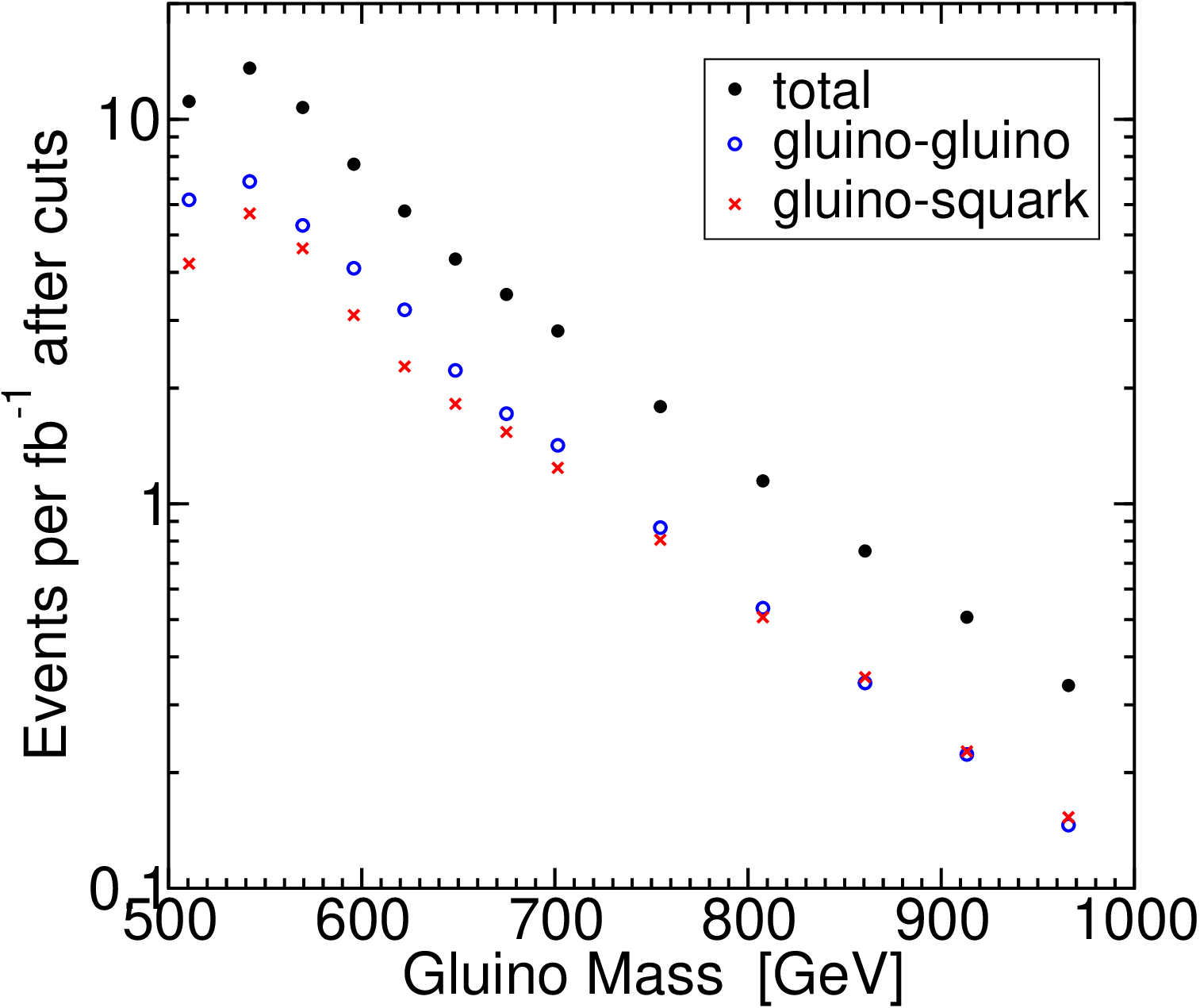}
\end{flushright}
\end{minipage}
\end{figure}

The cross-section after these cuts for LHC collisions with $\sqrt{s} = 14$
TeV is shown in figure \ref{fig:SSsig}, for points along the model line. 
Also shown are the two largest contributions, from $\tilde g\tilde g$
production and $\tilde g \tilde q$ production.%
The efficiency for the $\tilde g \tilde g$ part of the signal is about
$0.04$\%. Most of the $\tilde g \tilde q$ contribution to the signal is
due to production of a left-handed squark in association with a gluino,
since right-handed squarks usually decay directly to the LSP and the
corresponding quark. The cross-section after cuts is between 5 to 15 fb
for gluino masses less than 640 GeV, corresponding to the bulge region
where stop-mediated annihilation to top quarks dominates the dark matter
annihilation in the early universe. For comparison,
ref.~\cite{Kraml:2005kb} found backgrounds totaling less than 0.5 fb,
using very similar cuts (although a different event generation and
detector simulation). Therefore, strong evidence for this source of
supersymmetric events might be obtained with as little as a few fb$^{-1}$,
depending on the gluino mass, and to a lesser extent the squark masses.
This of course presumes that the backgrounds can be well understood, and
that wrong-sign assignments of lepton charges in e.g.~$t\overline t$
production are indeed not large and irreducible.  In the following, I will
optimistically assume this to be the case, and neglect backgrounds. 

Note that the cross-section after cuts is actually lower for the lowest
mass point in figure \ref{fig:SSsig} with $M_1 = 425$ GeV and $m_{\tilde
g} = 511$ GeV than for the next-higher mass point with $M_1 = 450$ GeV and
$m_{\tilde g} = 542$ GeV. This occurs for two main reasons. First, the
efficiency is lower for $M_1 = 425$ GeV because of the much smaller
stop-LSP mass difference, as noted in figure \ref{fig:massdiff}. Second,
the $M_1 = 450$ point has much larger branching fractions for right-handed
squarks to decay into gluinos, adding to the signal. 

[The contribution of right-handed squarks declines again for heavier
masses, and is eliminated by kinematics for points on the model line with
$m_{\tilde g}$ larger than about 600 GeV. The branching ratios for $\tilde
d_R \rightarrow d \tilde g$ and $\tilde u_R \rightarrow u \tilde g$ are
$(46\%, 38\%, 12\%)$ and $(21\%, 16\%, 5\%)$, respectively, for the model
line points with $M_1 = (450, 475, 500)$ GeV. For all other model line
points shown in figure \ref{fig:NLOsigma} and \ref{fig:SSsig}, the
branching ratios of $\tilde q_R \rightarrow q \tilde g$ are negligible.]

\begin{figure}[!tp]
\includegraphics[width=7.7cm,angle=0]{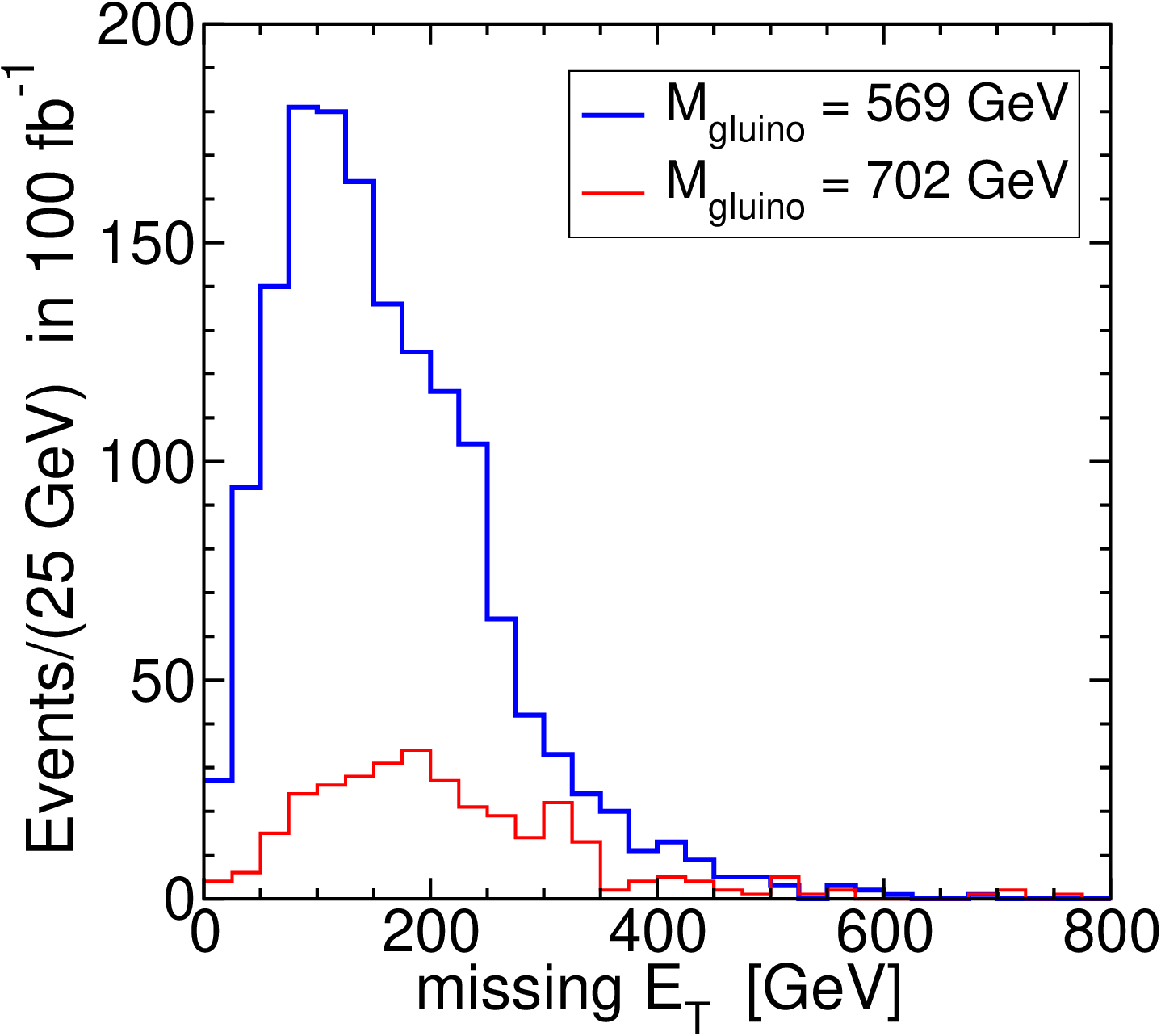}
\includegraphics[width=7.7cm,angle=0]{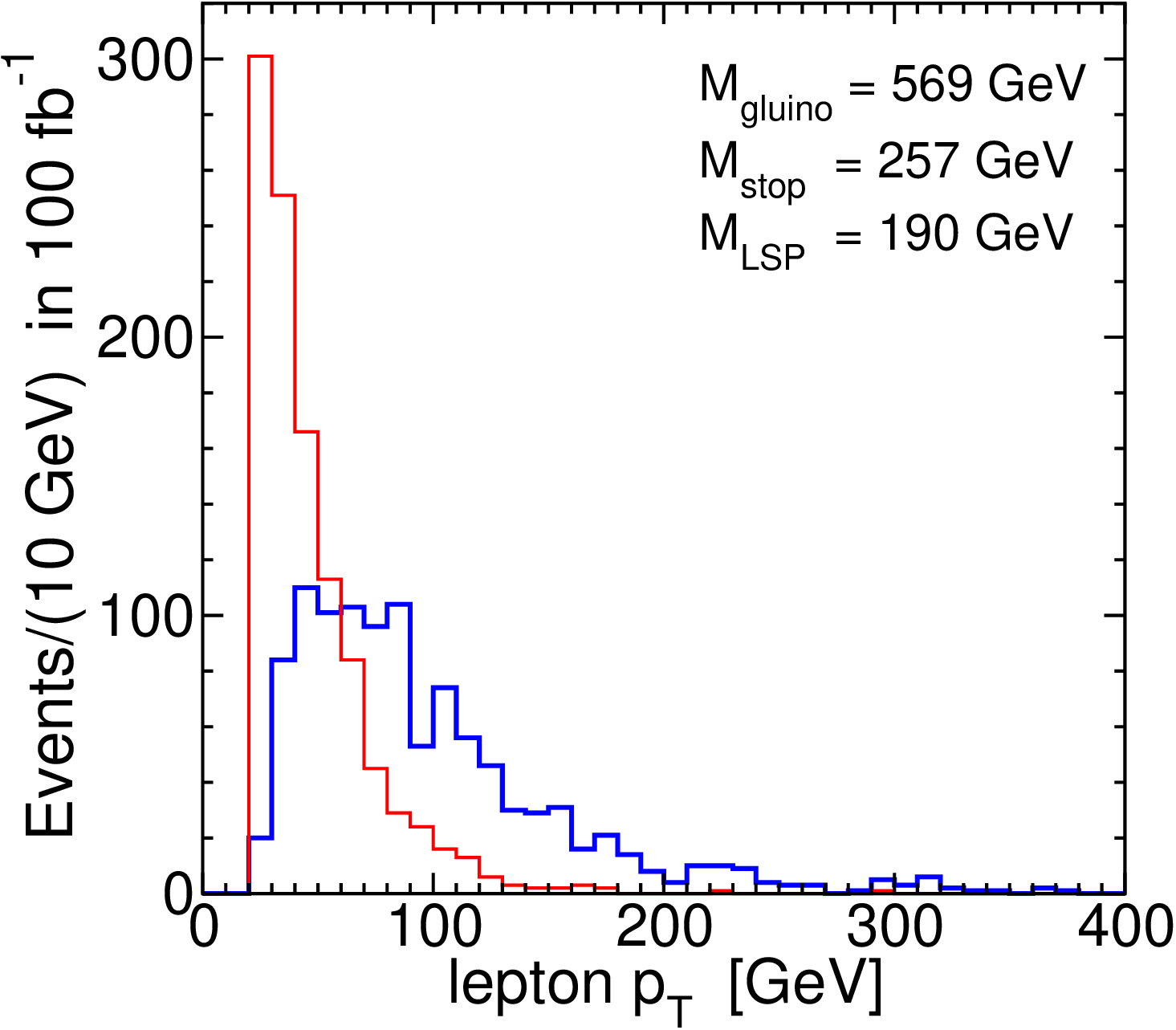}
\\
\includegraphics[width=7.7cm,angle=0]{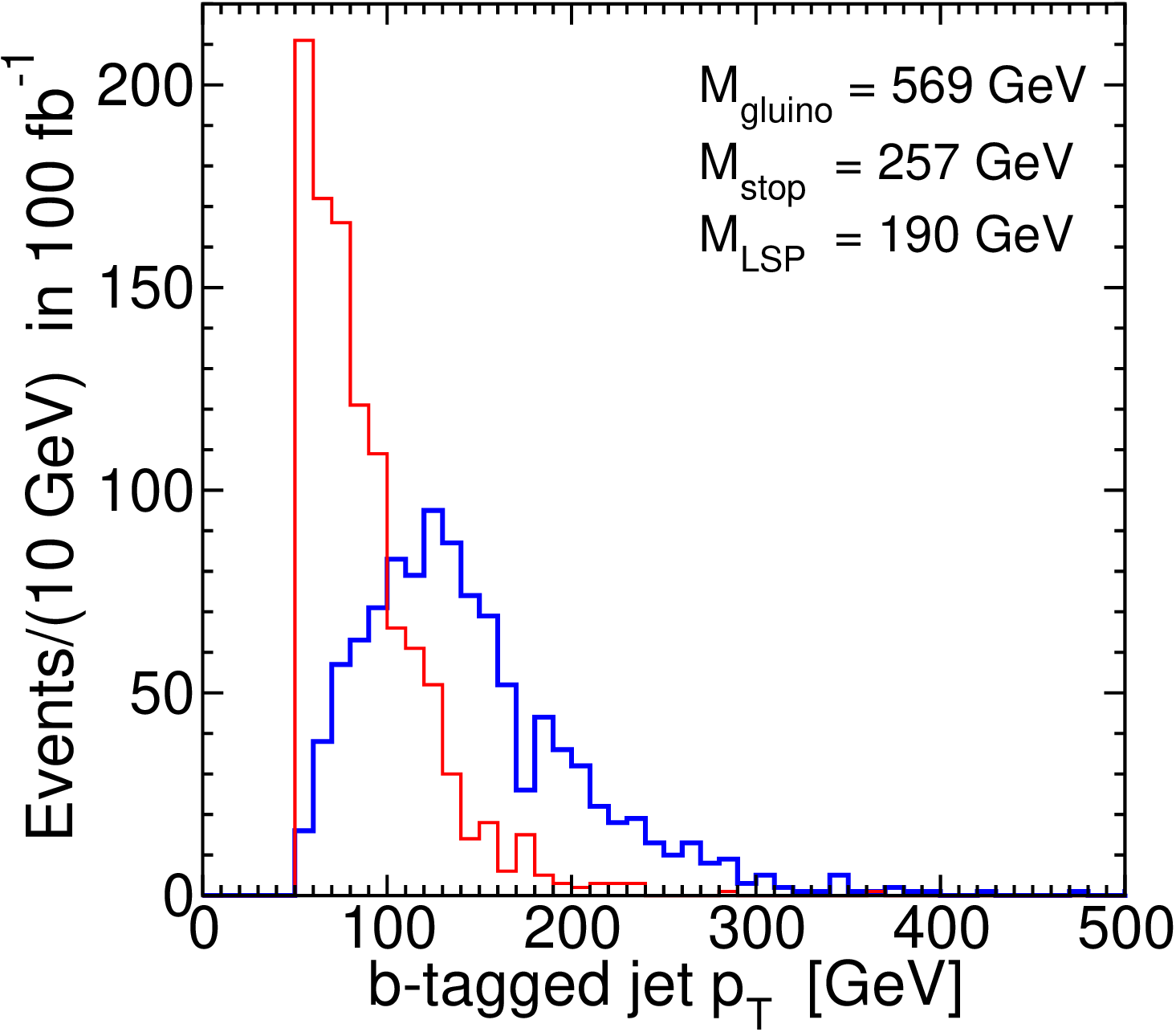}
\includegraphics[width=7.7cm,angle=0]{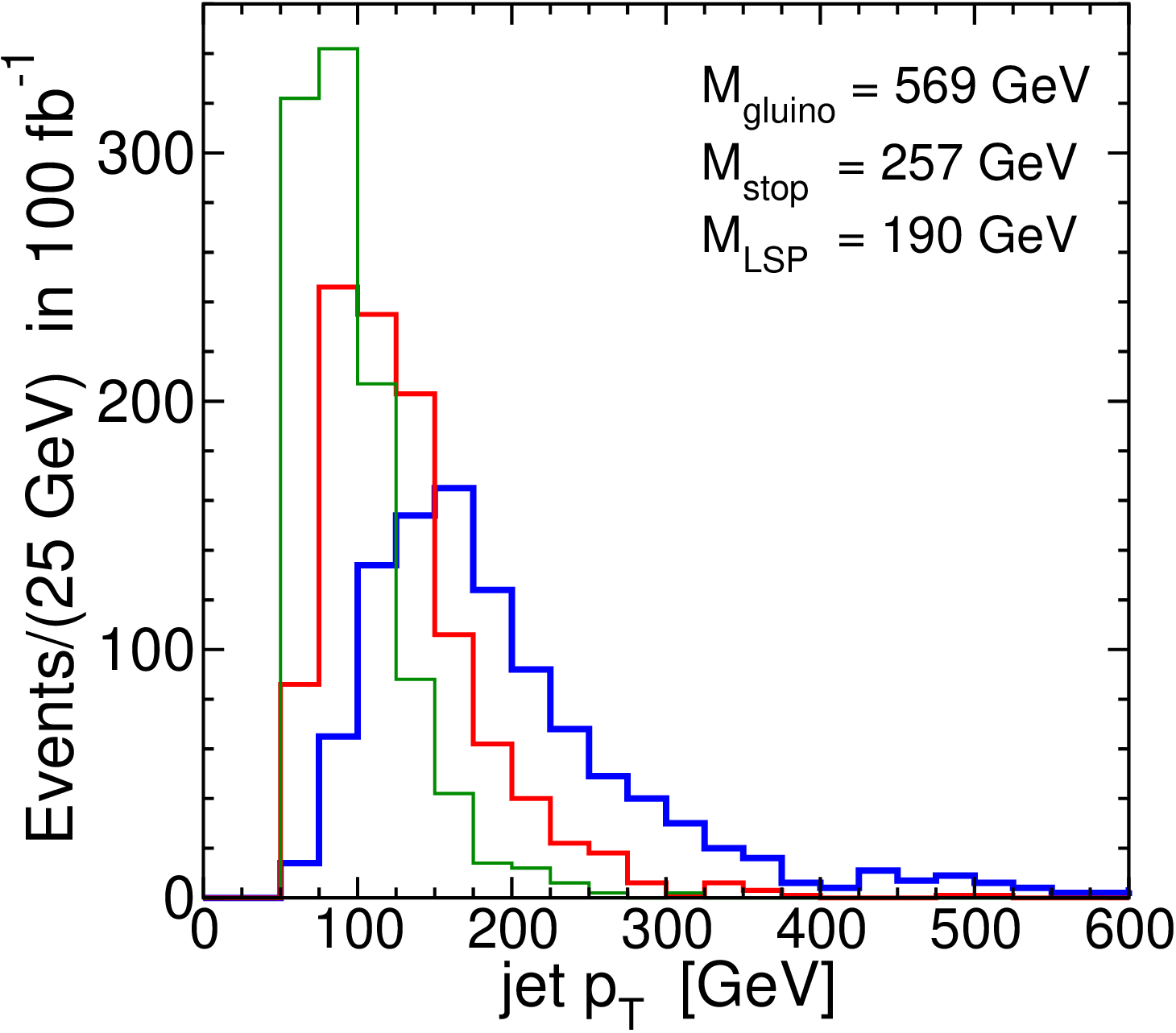}
\caption{\label{fig:pTdist}
Representative transverse momentum distributions for 
$\ell^\pm \ell^{\prime \pm} bbjj + \missET$ events 
from 100 fb$^{-1}$ of superpartner production, after the cuts described in
the text. The upper 
left panel shows the 
$\missET$
distribution (here without the $\missET$ cut) for two points on the model 
line with $m_{\tilde g} = 569$ 
and $702$ GeV.  The upper right 
panel shows the leading and subleading
lepton $p_T$ distributions for the model with $m_{\tilde g} = 569$,
and the lower panels show the leading and subleading $b$-jet 
distributions (left) and 
the three leading jet distributions (right) for the same model point.}
\end{figure}

To give an idea of the characteristics of the signal events, figure
\ref{fig:pTdist} shows the $\missET$, lepton $p_T$, and jet $p_T$ (with
and without $b$-tags) distributions for the events that pass the cuts,
given 100 fb$^{-1}$ of data for a representative model with $M_1 = 475$
GeV ($m_{\tilde g} = 569$ GeV). 
Clearly, raising the $\missET$ cut much farther above $100$ GeV would have
a significant unfortunate effect on the signal, even for heavier masses.
The same is true for the subleading lepton and jet $p_T$'s. On the other
hand, there is considerably more room to raise the cuts on the leading
lepton and jet $p_T$'s without a huge effect on the signal, should that
prove necessary to reduce backgrounds. Due to the practical difficulties
that are anticipated in commissioning $\missET$ at the LHC, it is also
tempting to consider dropping that cut altogether, since the same-sign
dileptons and jet cuts alone might be enough to distinguish the signal
from background. This may be, but figure \ref{fig:pTdist} shows that the
benefit accrued to the signal cross-section from relaxing the $\missET$
cut below 100 GeV is limited, especially for the critical case of models
with heavier gluinos, so for the purposes of the present analysis it will
be kept. 

\begin{figure}[!tp]
\begin{minipage}[]{0.37\linewidth}
\caption{\label{fig:btags}
The number of signal events after cuts in 100 fb$^{-1}$ 
with exactly 2, 3, 4 or 5 PGS heavy-flavor-tagged jets, for the point on the model
line described in section \ref{sec:modelline} 
with $M_1 = 500$ GeV, resulting in $m_{\tilde g} = 596$ GeV.
The breakdown into numbers of tight $b$ tags is also shown.
The relative frequencies of heavy-flavor tags
provides additional evidence for the $\tilde g \rightarrow t\tilde t_1$
and $\tilde t_1 \rightarrow c \tilde N_1$ interpretation of the signal.}
\end{minipage}
\begin{minipage}[]{0.62\linewidth}
\includegraphics[width=7.0cm,angle=0]{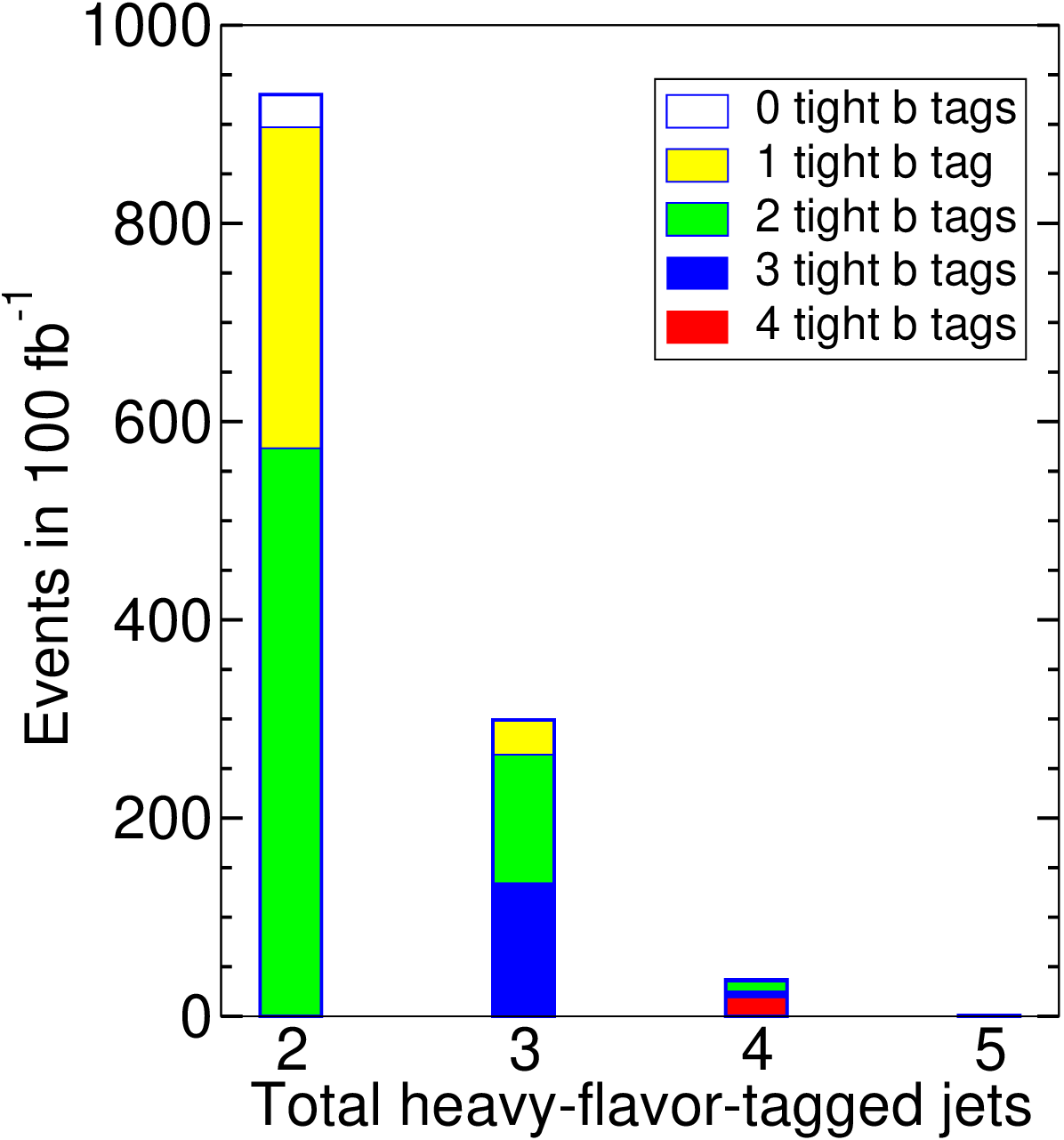}
\end{minipage}
\end{figure}

The frequency of heavy-flavor-tagged jets in the signal sample can help 
to confirm that the signal is really due to gluino pairs decaying to 
stops that in turn decay to charm quarks and LSPs. 
The number of events with 2, 3, 4 or 5  heavy-flavor-tagged jets is shown in
figure \ref{fig:btags} for the point on the model line with 
$M_1 = 500$ GeV, for 100 fb$^{-1}$. Also included is the breakdown of
these events into tight $b$-tags as reported by PGS, with efficiencies
for central high-$p_T$ jets of approximately 40\% for true $b$-jets,
9\% for $c$-jets, and $0.1$\% for $g,u,d,s$ jets.
As a simple-minded check, 
one can assume that the $m(b\ell) < 160$ GeV requirement
preselects only events that have the true $b$-jets tagged, so that 
additional heavy-flavor tags come from the true charm jets, and the
numbers of events with 2, 3, 4, and 5 heavy flavor tags 
should be roughly in the proportion
$n_2:n_3:n_4:n_5 = (1 - P_c)^2 : 2 P_c (1 - P_c): P_c^2:0$, where $P_c$ is the
probability of a true charm jet to get a heavy flavor tag. Using $n_2 = 930$
for the example in figure \ref{fig:btags}, and $P_c = 0.13$, one would
predict $n_3 = 278$ and $n_4 = 21$ and $n_5 = 0$, 
in not unreasonable agreement for such a naive estimate with 
the actual finding of $n_3 = 299$ and $n_4 = 37$ and $n_5 = 1$. 
This information can be used to clearly 
distinguish the scenario under study here from similar ones in which
the stop-LSP mass difference is large enough to allow
$\tilde t_1 \rightarrow b W \tilde N_1$, or where strict
minimal flavor violation leads to a dominant or competitive
4-body decay $\tilde t_1 \rightarrow b f \bar f' \tilde N_1$,
either of
which would lead instead to a parton-level same-sign dilepton signature of
\beq
pp \rightarrow \ell^\pm\ell^{\prime \pm} bbbbjjjj + \missET.
\eeq
In this case, one would clearly expect many more events with 3, 4 and even
5 or more (since half of the hadronic $W$ decays will result in a true
charm jet) heavy-flavor tags relative to the number with 2 tags, compared
to the situation in figure \ref{fig:btags}. Measuring the numbers of
heavy-flavor tags within the $\ell^\pm\ell^{\prime \pm} bbjj+\missET$
signal sample can therefore establish whether $m_{\tilde t_1} - m_{\tilde
N_1} < 85$ GeV. Of course, the specific numbers for heavy-flavor tagging
in the actual LHC detector environments might be quite different from
those assumed here, but the principle should still apply. 

In the remainder of this paper, I will examine some strategies for
obtaining information about the gluino, squark and LSP mass spectrum. Note
that variables like $m_{T2}$ \cite{mTtwoA,mTtwoB} are hampered, for the
same-sign dilepton event topology, by the inevitable presence of two
neutrinos with unknown momenta in addition to the two LSPs in each event.
I have therefore not attempted the difficult task of seeing whether this
can give useful information when 
applied to the same-sign lepton sample. 
The definite absence of dilepton mass edges
eliminates another commonly used tool
\cite{dileptons}-\cite{Lester:2005je} for reconstructing superpartner
decay chains. Instead, I will consider mass estimators that use
scalar-summed transverse momenta and single-lepton mass edges from visible
gluino decay products. 

\section{Mass estimators from scalar-summed transverse momenta}
\label{sec:ptsums}
\setcounter{equation}{0}
\setcounter{footnote}{1}

One of the most important efforts in a future LHC analysis of
supersymmetry will be to obtain measurements, or at least estimates, of
the superpartner masses. The purpose of this section is to consider
observables that can serve as estimators of the masses of the
superpartners produced, using scalar sums of the lepton and jet $p_T$'s.
There has been considerable effort in this area, often using the
observables $H_T$ and $M_{\rm eff}$ for events with jets and $\missET$.
Here, I will study the prospects for using similar observables, but in the
hopefully cleaner context of the $\ell^\pm\ell^{\prime \pm} bb jj +
\missET$ signal discussed in the previous section. To this end, consider
four mass estimators defined by: 
\beq
M_A &=& \sum_n p_T(j_n) + \sum_{n=1,2} p_T(\ell_n) + \missingET,
\label{eq:defMA}
\\
M_B &=& \sum_{n=1,2,3,4} p_T(j_n) + \sum_{n=1,2} p_T(\ell_n) + \missingET,
\label{eq:defMB}
\\
M_C &=& \sum_n p_T(j_n) + \sum_{n=1,2} p_T(\ell_n) ,
\label{eq:defMC}
\\
M_D &=& \sum_{n=1,2,3,4} p_T(j_n) + \sum_{n=1,2} p_T(\ell_n),
\label{eq:defMD}
\eeq
where the jet labels are ordered by $p_T(j_1) > p_T(j_2) > p_T(j_3) >
\ldots$. As usual, the idea is that the transverse momenta of the decay
products should be approximately linear in the mass of the pair of heavy
particles produced. The first observable, $M_A$, simply sums over all
visible object $p_T$'s and the $\missET$. The second observable $M_B$ is
motivated by the ideas that the sum over only the leading 4 jets should be
less sensitive to theoretical uncertainties due to extra jets from the
underlying event, and that the signal includes at least four quark
partons.  The other two observables $M_{C,D}$ are the same as $M_{A,B}$
except that $\missET$ is not included. This is motivated by the fact that
$\missET$ may be particularly difficult to obtain accurately, especially
in early running of the LHC. 

\begin{figure}[!tp]
\includegraphics[width=8.0cm,angle=0]{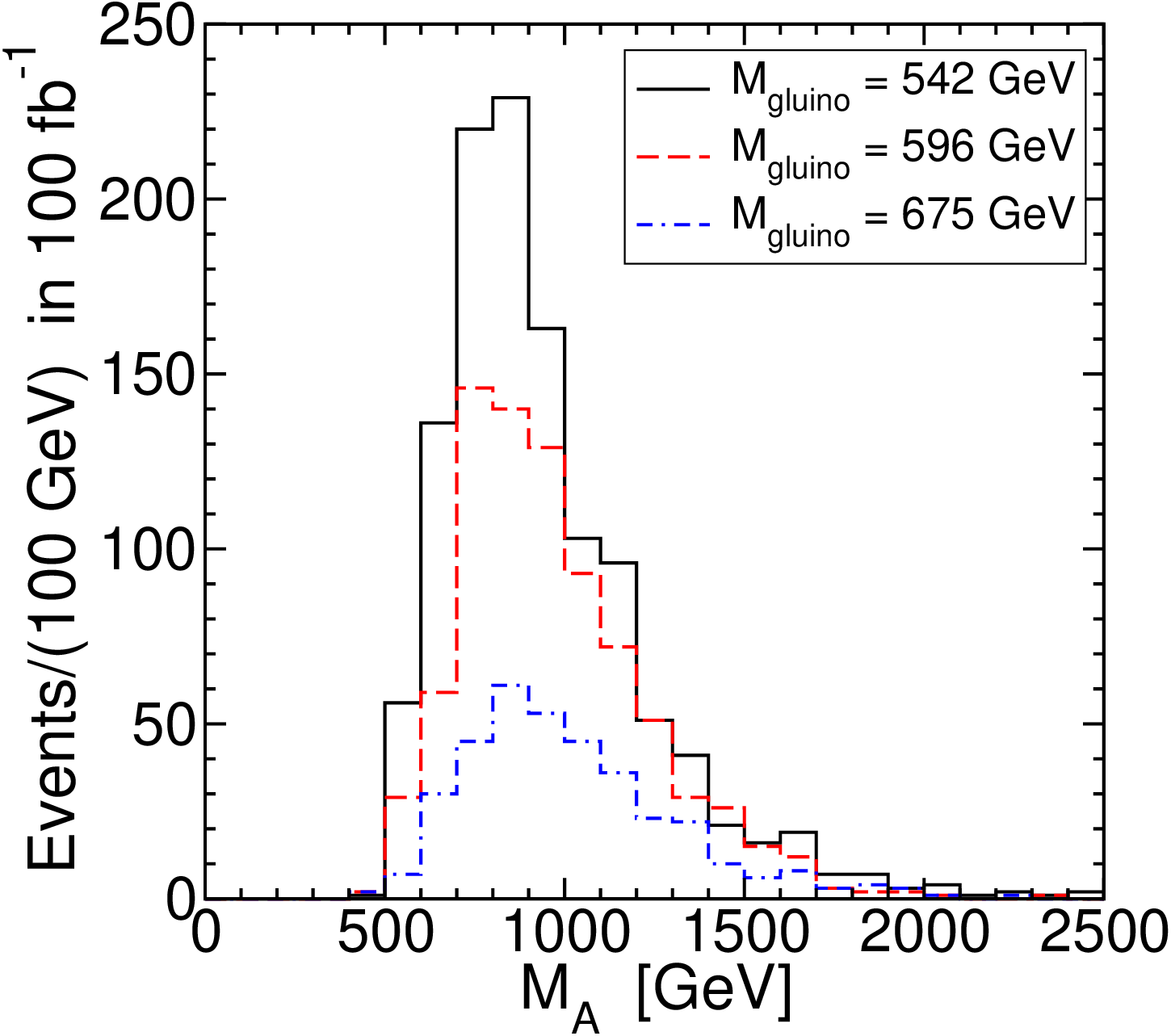}
\includegraphics[width=8.0cm,angle=0]{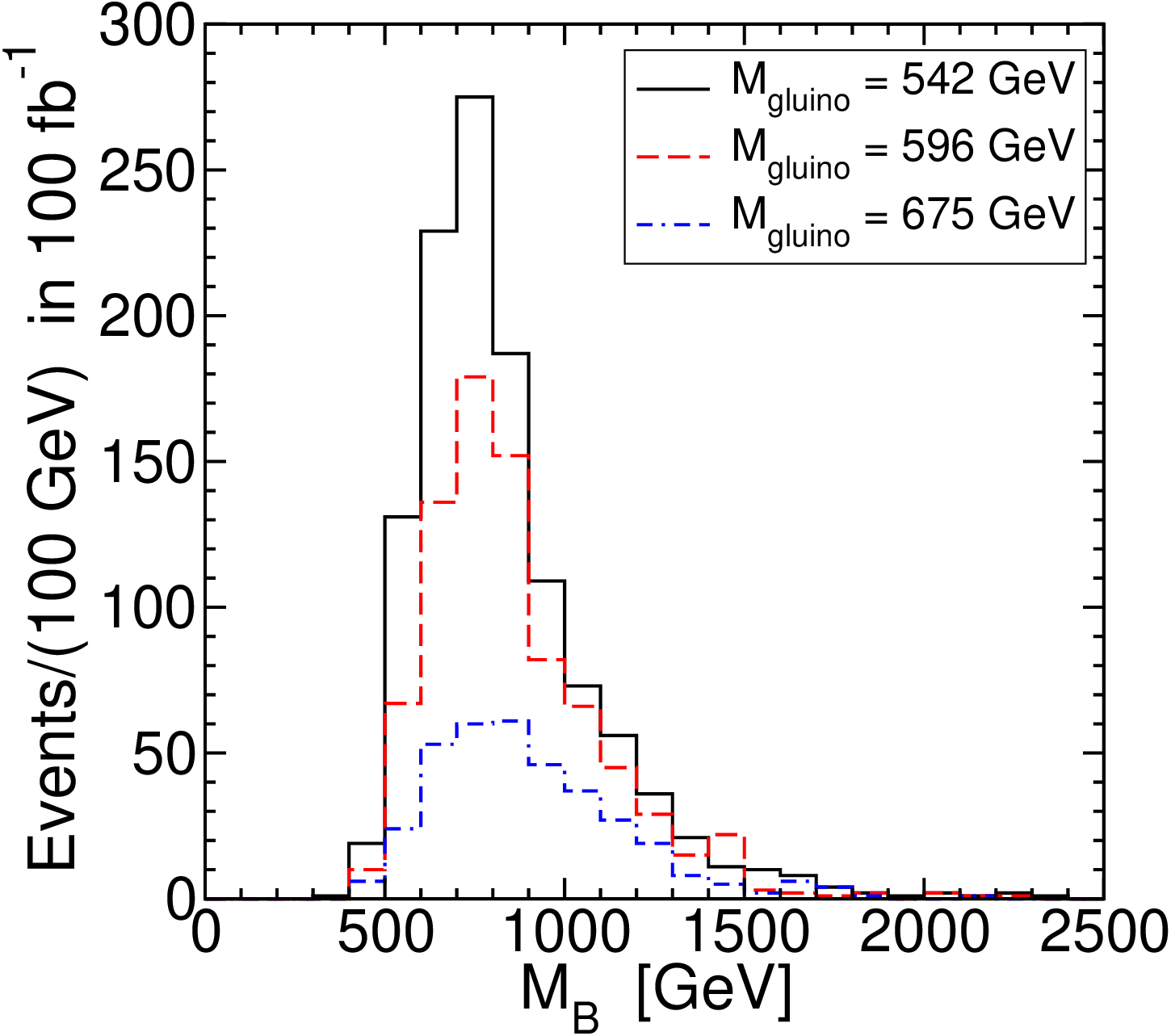}
\\
\includegraphics[width=8.0cm,angle=0]{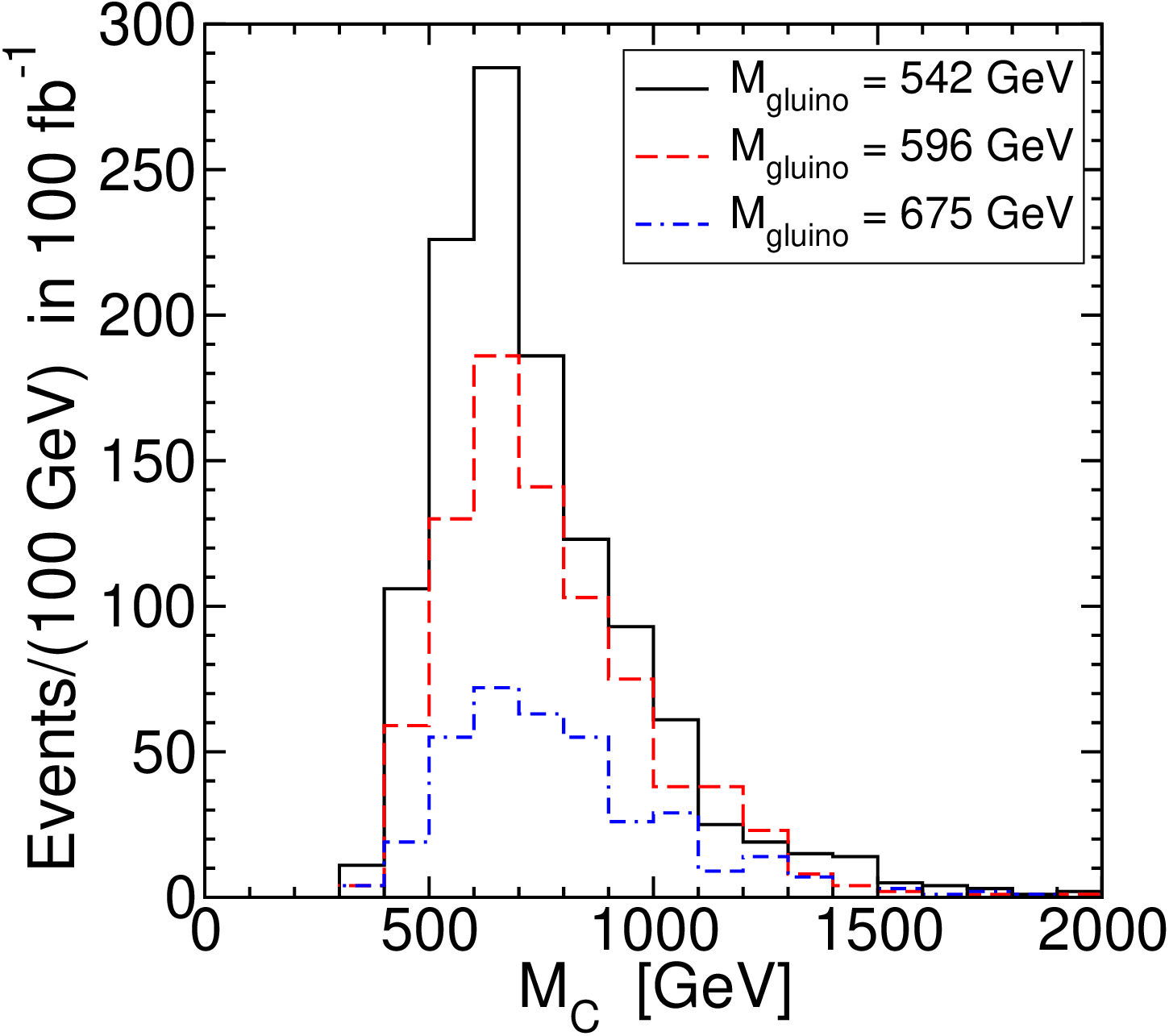}
\includegraphics[width=8.0cm,angle=0]{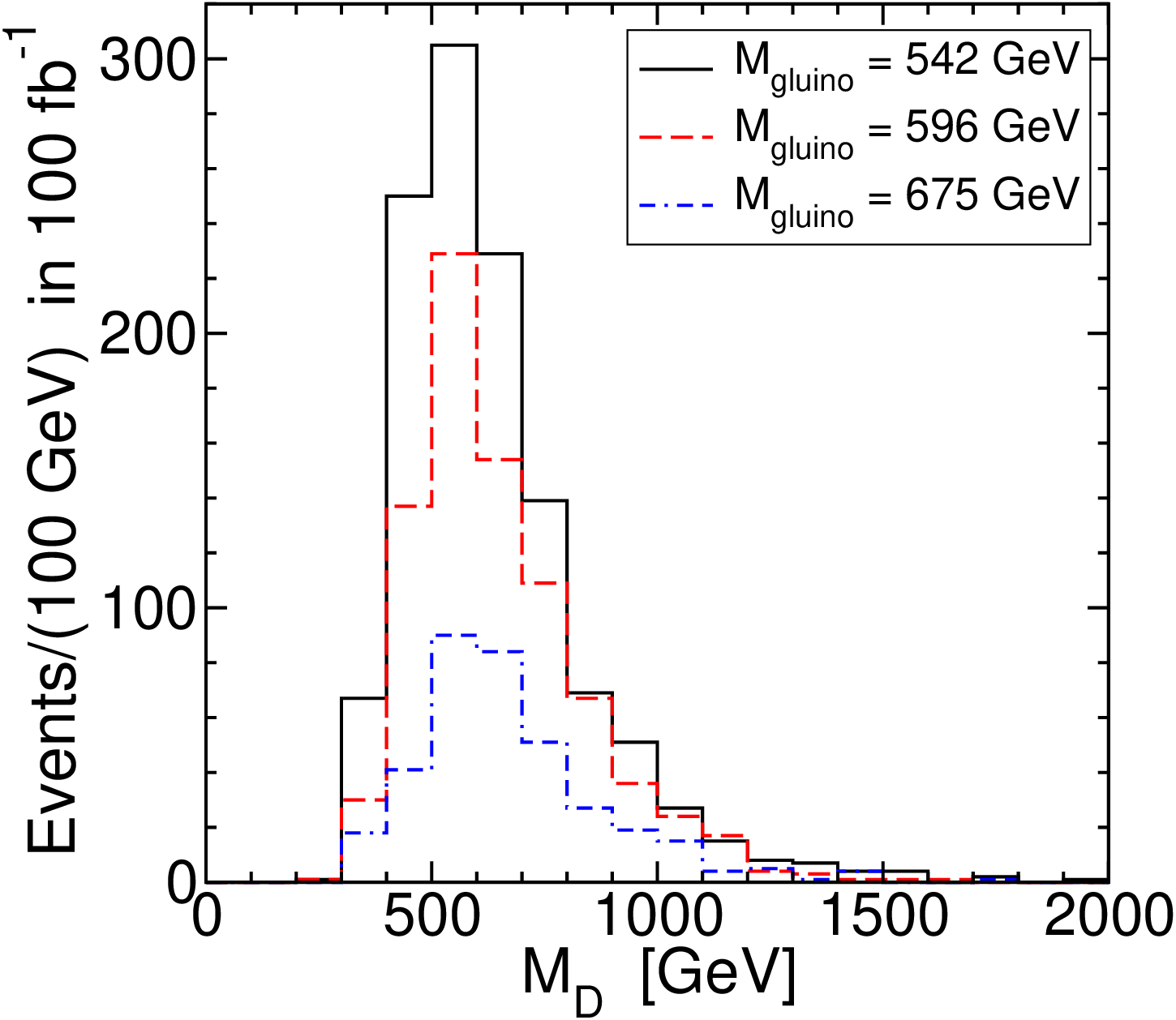}
\caption{\label{fig:MABCDdist}
Distributions for the mass estimators $M_{A,B,C,D}$ defined
by eqs.~(\ref{eq:defMA})-(\ref{eq:defMD}), 
for 100 fb$^{-1}$ of events for
three models along the model line 
described in section \ref{sec:modelline}, with $M_1 = 450$, 500, and 575
GeV.}
\end{figure}

Using the same event selection criterion as in the previous section,
the distributions for these four mass estimators are shown in figure
\ref{fig:MABCDdist}, for 100 fb$^{-1}$ of data with 
three representative points on the model line defined in
section \ref{sec:modelline}
with $m_{\tilde g} = 542$, 596, and 675 GeV.
Even from these coarse-binned distributions, it is apparent that the
shapes of the distributions are distinguishable from each other.

To determine a sharper empirical relation between these mass estimators
and the superpartner mass scale, I performed an unbinned maximum
likelihood fit to 100 fb$^{-1}$ of generated events for each of 9 model
points. Because the distributions of $M_X$ (for $X = A,B,C,D$) are clearly
far from Gaussian, better results are obtained by fitting them instead to
the class of functions known as generalized inverse Gaussian distributions
(with $x = M_X$): 
\beq
f(x) \,=\, \frac{1}{n} (x - x_0)^{-c} \,{\rm exp} \left [
-\frac{b (x - x_0 - a)^2}{2 (x-x_0)} \right ].
\eeq
Here $a$, $b$, and $c$ are the fit parameters, and $x_0$ is the 
minimum of the distribution following simply from 
the jet and lepton $p_T$ and $\missingET$ cuts. 
(In the present analysis, $x_0$ is equal to 325 GeV for $X=A,B$ and  
225 GeV for $X = C,D$.) The normalization condition 
\beq
\int_{x_0}^\infty f(x) \, dx = 1
\eeq
implies that
\beq
\ln(n) = a b + (1-c) \ln(a) + \ln[2 K_{c-1}(a b)],
\eeq
with $K_i(z)$ the modified Bessel function of the second kind.
The peak of each distribution, defined as the value where $df/dx=0$, 
is then obtained as
\beq
M_X^{\rm peak} = x_0 + (\sqrt{a^2 b^2 + c^2} - c)/b,
\eeq
where $a,b,c$ are set equal to their best-fit values\footnote{The best
fits obtained in the following almost always turn out to have $c$ very
close to $3/2$, corresponding to the special case known as an ordinary
inverse Gaussian distribution (not to be confused with a normal Gaussian
distribution).  I do not know the explanation for this.} after maximizing
the log likelihood function. 

\begin{figure}[!tp] 
\includegraphics[width=8.9cm,angle=0]{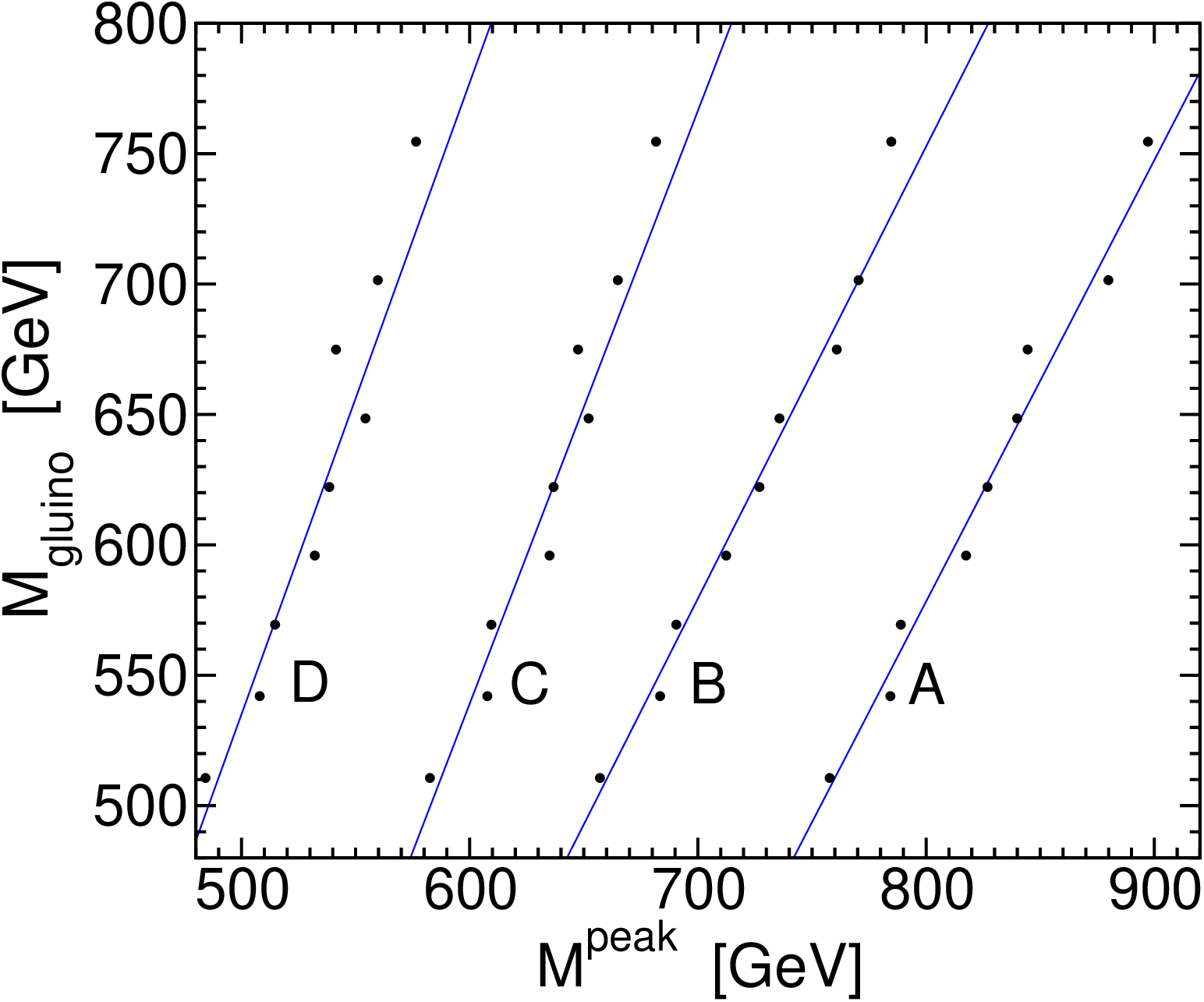}
\caption{\label{fig:MABCDgluino} The gluino mass is compared to the fitted
peak values of the distributions of the mass estimators $M_{A,B,C,D}$
defined by eqs.~(\ref{eq:defMA})-(\ref{eq:defMD}). Results for 100
fb$^{-1}$ of events for each of 9 individual models along the model line
are shown as black dots, together with the best fit lines
eqs.~(\ref{eq:linearA})-(\ref{eq:linearD}).} \end{figure}

Performing a linear regression (with the variance for each model 
taken inversely proportional to the number of events found
in 100 fb$^{-1}$) gives  relationships between the mass estimators 
$M_X^{\rm peak}$ and the gluino mass:
\beq
M_{\tilde g} 
&=& 1.693 M_A^{\rm peak} - 776\>{\rm GeV},
\label{eq:linearA}
\\
&=& 1.733 M_B^{\rm peak} - 634\>{\rm GeV},
\label{eq:linearB}
\\
&=& 2.274 M_C^{\rm peak} - 825\>{\rm GeV},
\label{eq:linearC}
\\
&=& 2.422 M_D^{\rm peak} - 676\>{\rm GeV}.
\label{eq:linearD}
\eeq
The comparison between the linear fits and the values obtained for the
individual models are shown in figure \ref{fig:MABCDgluino}.%
 The smaller
slopes of the $M_{\tilde g}$ vs.~$M_{A,B}^{\rm peak}$ lines would seem to
make them more useful as mass estimators than $M_{C,D}^{\rm peak}$,
although this depends crucially on the presently unknown quality of the
$\missET$ determination. 

In general, the mass estimators might be expected to be roughly
proportional to $M_{\rm eff} = M_{\rm SUSY} - m_{\tilde N_1}^2/M_{\rm
SUSY}$, where $M_{\rm SUSY}$ is a signal cross-section weighted average of
the superpartner masses, in this case the gluino and left-handed squark
masses.  In models with a slightly larger $m_{\tilde q_R} - m_{\tilde g}$
mass difference, the decay $\tilde q_R \rightarrow q \tilde g$ would be
more important, and $m_{\tilde q_R}$ would be weighted more strongly into
$M_{\rm SUSY}$.  In the model line under study, the LSP mass and the
gluino mass are very nearly proportional, and the squark masses are also
tightly correlated with the gluino mass. In general, since the presence of
the signal depends crucially on the Majorana nature of the gluino, this
method should be useful to obtain a rough estimate of the gluino mass,
albeit with some mild model assumptions. 

\section{Endpoints of visible gluino decay products}
\label{sec:endpoints}
\setcounter{equation}{0}
\setcounter{footnote}{1}

Another method that can be used to gain information about the superpartner
masses is to look at the invariant mass distributions of identified
visible products of the gluino decay. This method has already been
extensively studied in ref.~\cite{Kraml:2005kb}, in a situation similar to
the present one, but with a relatively much lighter top squark and LSP,
and taking the squarks to be much heavier. In the model scenario under
study here, the presence of a large component of $\tilde g \tilde q
\rightarrow \tilde g \tilde g q$ in the signal causes a significant
additional source for confusion in identifying the jets following from the
gluino decay. 

\begin{figure}[!tp]
\begin{minipage}[]{0.37\linewidth}
\caption{\label{fig:massdist}
The parton-level predictions for the
distributions of the invariant masses $m(b\ell c)$, $m(bc)$, and 
$m(\ell c)$,
for the decay $\tilde g \rightarrow t \tilde t_1$ followed by
$\tilde t_1 \rightarrow c \tilde N_1$ and $t \rightarrow bW$ and
$W \rightarrow \ell\nu$. The masses used are 
$m_{\tilde g} = 596$ GeV, $m_{\tilde t_1} = 261$
GeV, and $m_{\tilde N_1} = 201$ GeV, corresponding to a point on the model
line with $M_1 = 500$ GeV. 
The endpoints of the distributions
are at $m(b\ell c) = 354$ GeV and $m(bc) = 283$ GeV and $m(\ell c) = 320$ GeV.}
\end{minipage}
\begin{minipage}[]{0.6\linewidth}
\begin{flushright}
\includegraphics[width=9.4cm,angle=0]{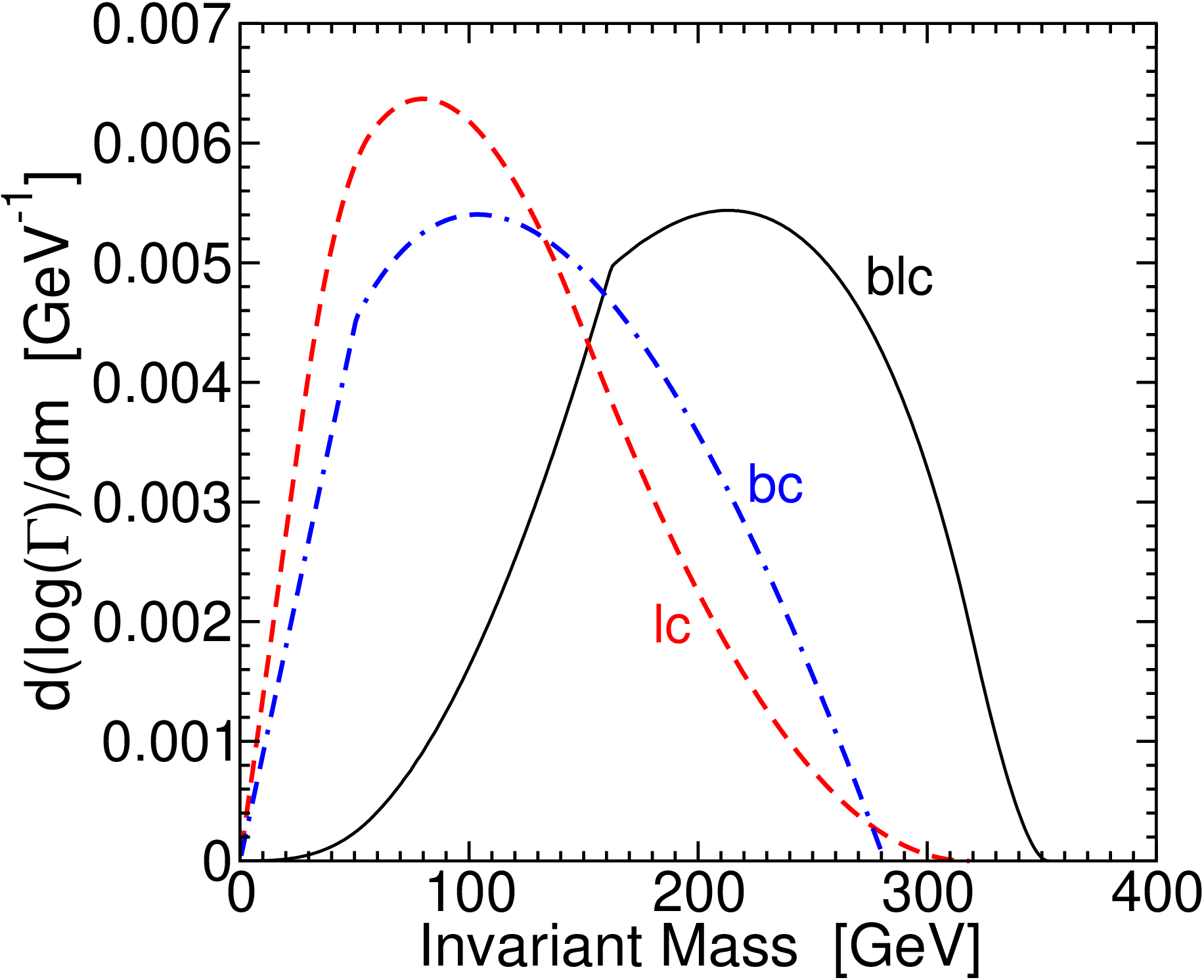}
\end{flushright}
\end{minipage}
\end{figure}

The parton-level kinematic endpoints from the decay 
$\tilde g \rightarrow t \tilde 
t_1$ followed by $\tilde t_1 \rightarrow c \tilde N_1$ and $t 
\rightarrow b \ell \nu$ are \cite{Kraml:2005kb}: 
\beq
m^2(\ell c)_{\rm max} &=& \frac{1}{2} 
\bigl ( 1 - m^2_{\tilde N_1}/m^2_{\tilde t_1} \bigr )
\left [ m^2_{\tilde g} - m_{\tilde t_1}^2 - m_t^2 + 
\lambda^{1/2} (m_{\tilde g}^2, m_{\tilde t_1}^2, m_t^2) \right ],
\label{eq:mlcmax}
\\
m^2(bc)_{\rm max} &=& \bigl (1 - m_W^2/m_t^2 \bigr ) m^2(\ell c)_{\rm max},
\label{eq:mbcmax}
\\
m^2(b\ell c)_{\rm max} &=&  m^2(\ell c)_{\rm max} + m_t^2 - m_W^2,
\label{eq:mblcmax}
\eeq
where $\lambda(x,y,z) = x^2 + y^2 + z^2 - 2 x y - 2 x z - 2 y z$. Note
that these endpoints are not independent; knowing any one of them yields
the others, given the known masses of the top quark and $W$ boson. (The
widths of the particles, and the mass of the bottom quark, are neglected
here.) The corresponding distributions \cite{Kraml:2005kb} for the model
depicted in figure \ref{fig:spectrum500} with $m_{\tilde g} = 596$ GeV,
$m_{\tilde t_1} = 260.5$ GeV, and $m_{\tilde N_1} = 200.8$ GeV are shown
in figure \ref{fig:massdist}. 
Ref.~\cite{Kraml:2005kb} performed fits to the shapes of $bc$ and $\ell c$
mass distributions, finding that the quality of the fits was made worse by
leptons from taus in the top decays, among other effects. In the present
case, there is a serious additional (and in practice, unknown) effect on
the shape from wrong jet assignments due to the presence in some events of
$\tilde q_L \rightarrow q \tilde g$, and to a lesser extent $\tilde q_R
\rightarrow q \tilde g$. Therefore, it is probably more robust to
concentrate on the endpoints of the distributions.  From figure
\ref{fig:massdist} one sees that the $m(\ell c)$ distribution is extremely
shallow near the endpoint, making it very difficult to determine the
endpoint from data.  The $m(bc)$ and $m(b\ell c)$ distributions are much
steeper near their respective endpoints, so I will only consider them. It
should be noted that different events contribute to the near-endpoint
regions of these two distributions, even though the positions of the
endpoints are algebraically related. 

To mitigate the problem of wrong jet assignments, I use a subset of events
selected by the procedure described in section \ref{sec:sstops}, with the
additional constraint that the pairing of the two leptons with $b$-tagged
jets consistent with top decays [$m(b\ell) < 160$ GeV] is {\em unique}.
(This reduces the signal efficiency by about a factor of 3.) For each
$b\ell$ pair, the putative charm jet is taken to be the one with the
smallest value of $m(b\ell c)$ selected from among those with $p_T > 35$
GeV. This selection means that far below the endpoints, there may well be
many wrong assignments (both from extra jets in the underlying event and
from jets produced in squark decays being assigned to the charm jet role),
but near the endpoints the assignments are made correctly with greater
frequency. 
\begin{figure}[!tp]
\includegraphics[width=7.93cm,angle=0]{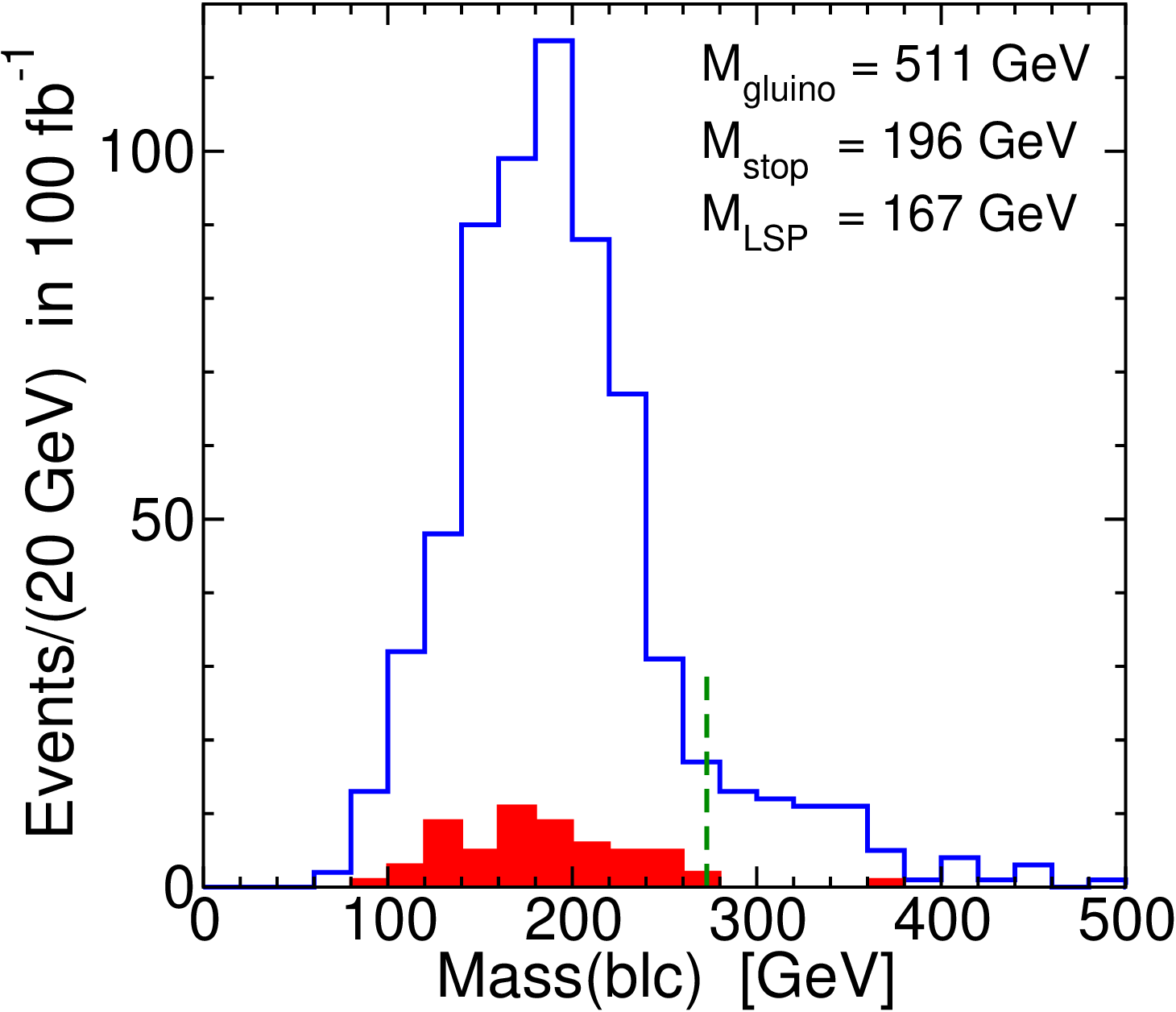}~~~
\includegraphics[width=7.93cm,angle=0]{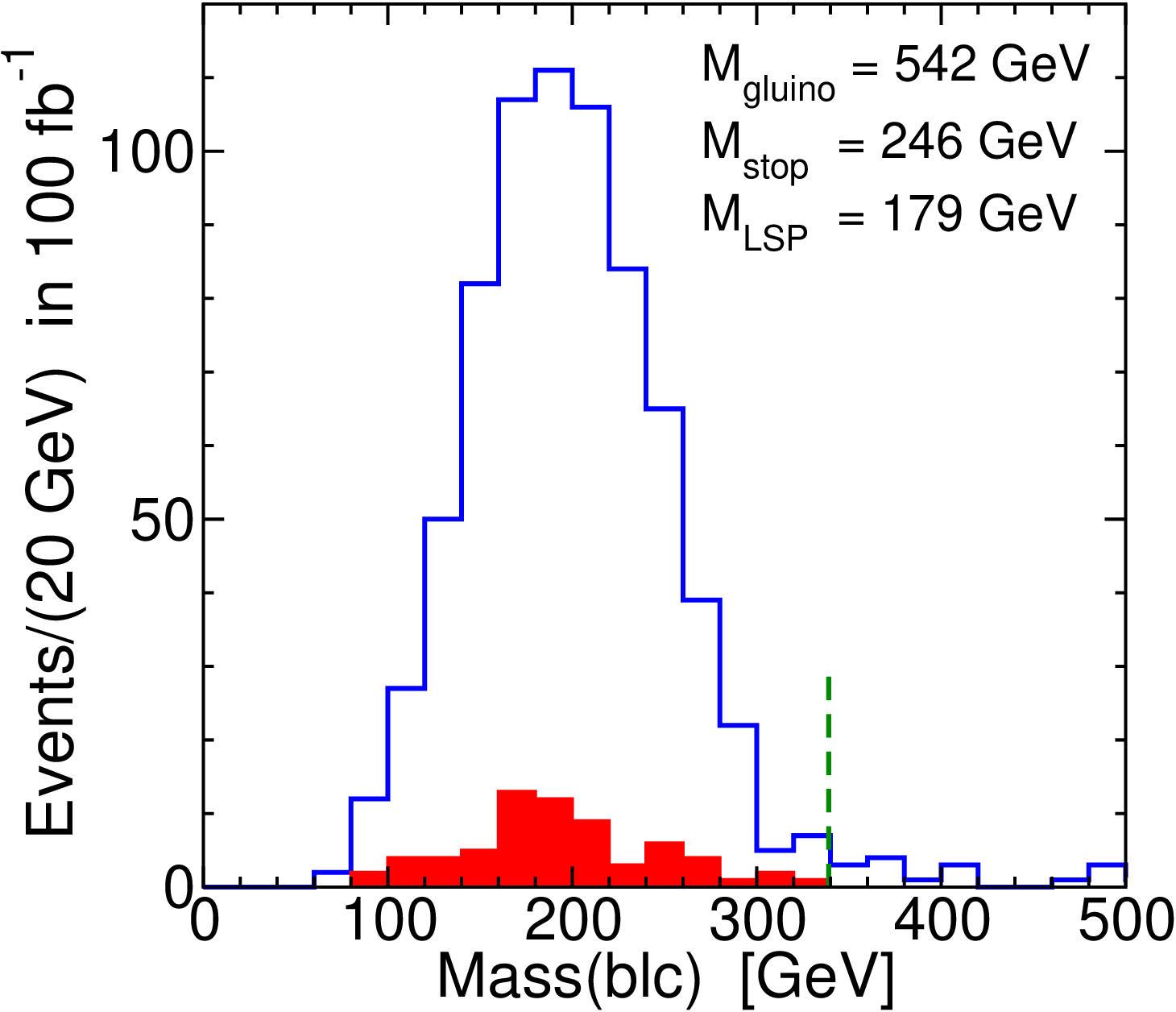}
\\
\includegraphics[width=7.93cm,angle=0]{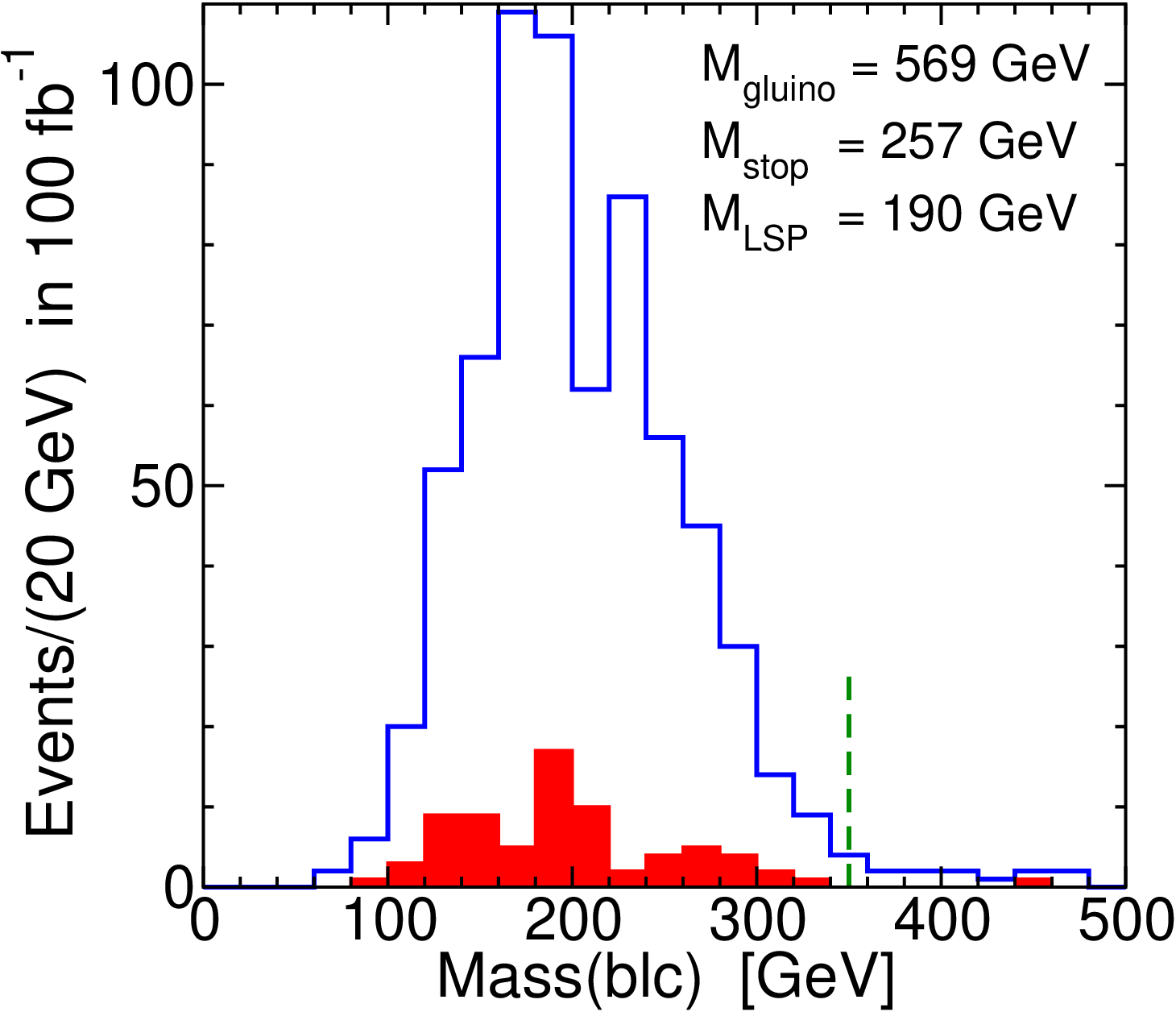}~~~
\includegraphics[width=7.93cm,angle=0]{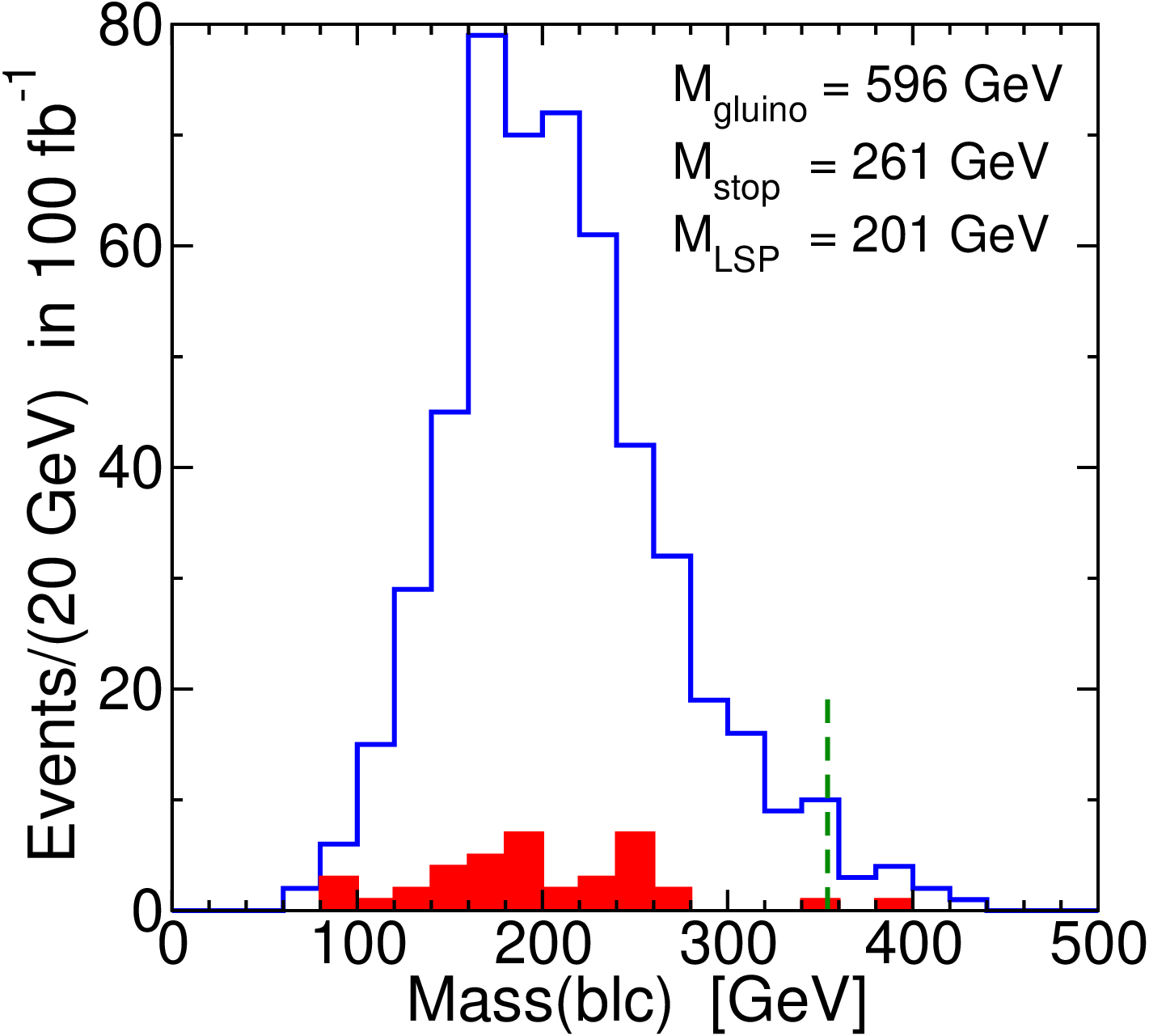}
\\
\includegraphics[width=7.93cm,angle=0]{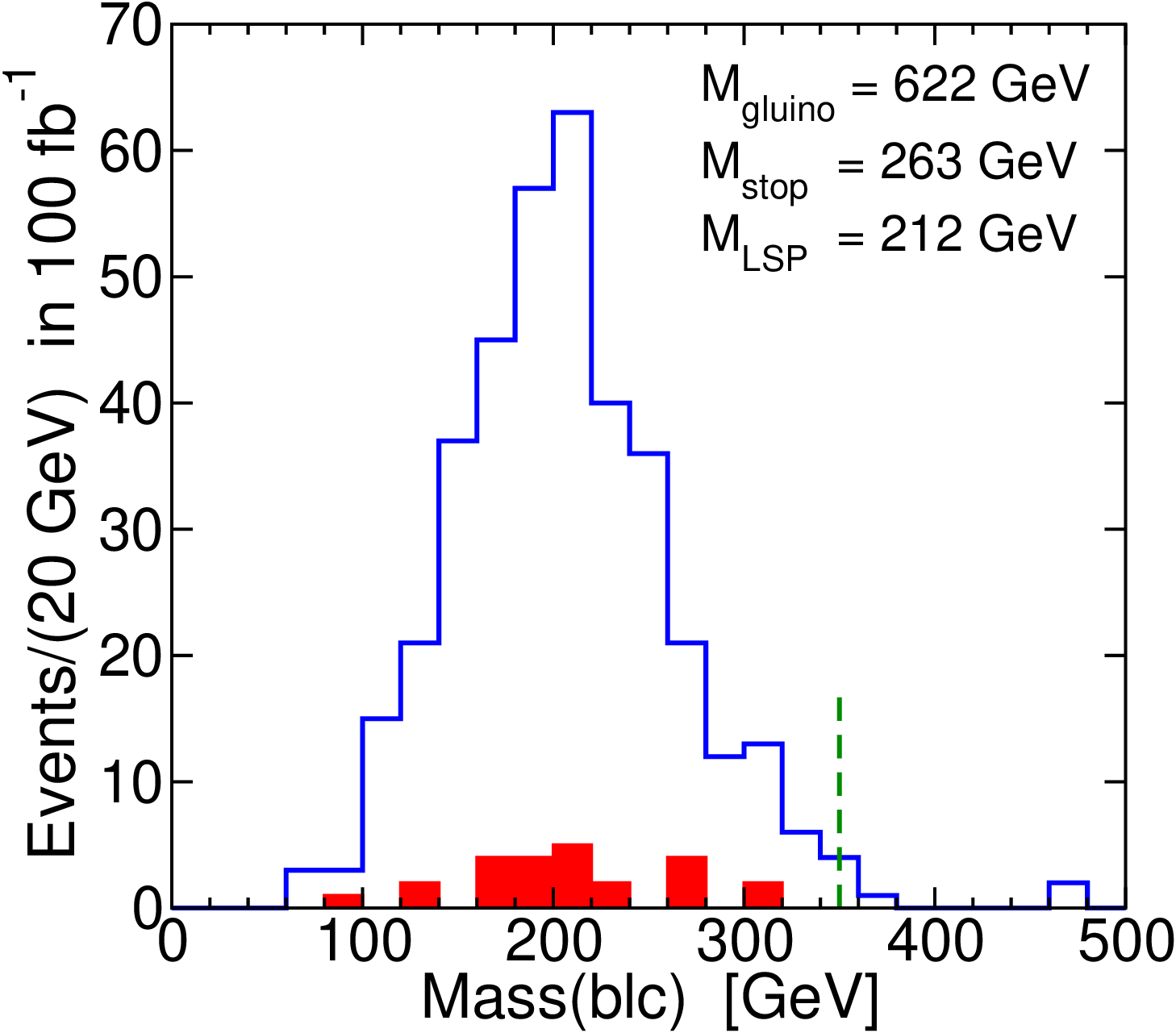}~~~
\includegraphics[width=7.93cm,angle=0]{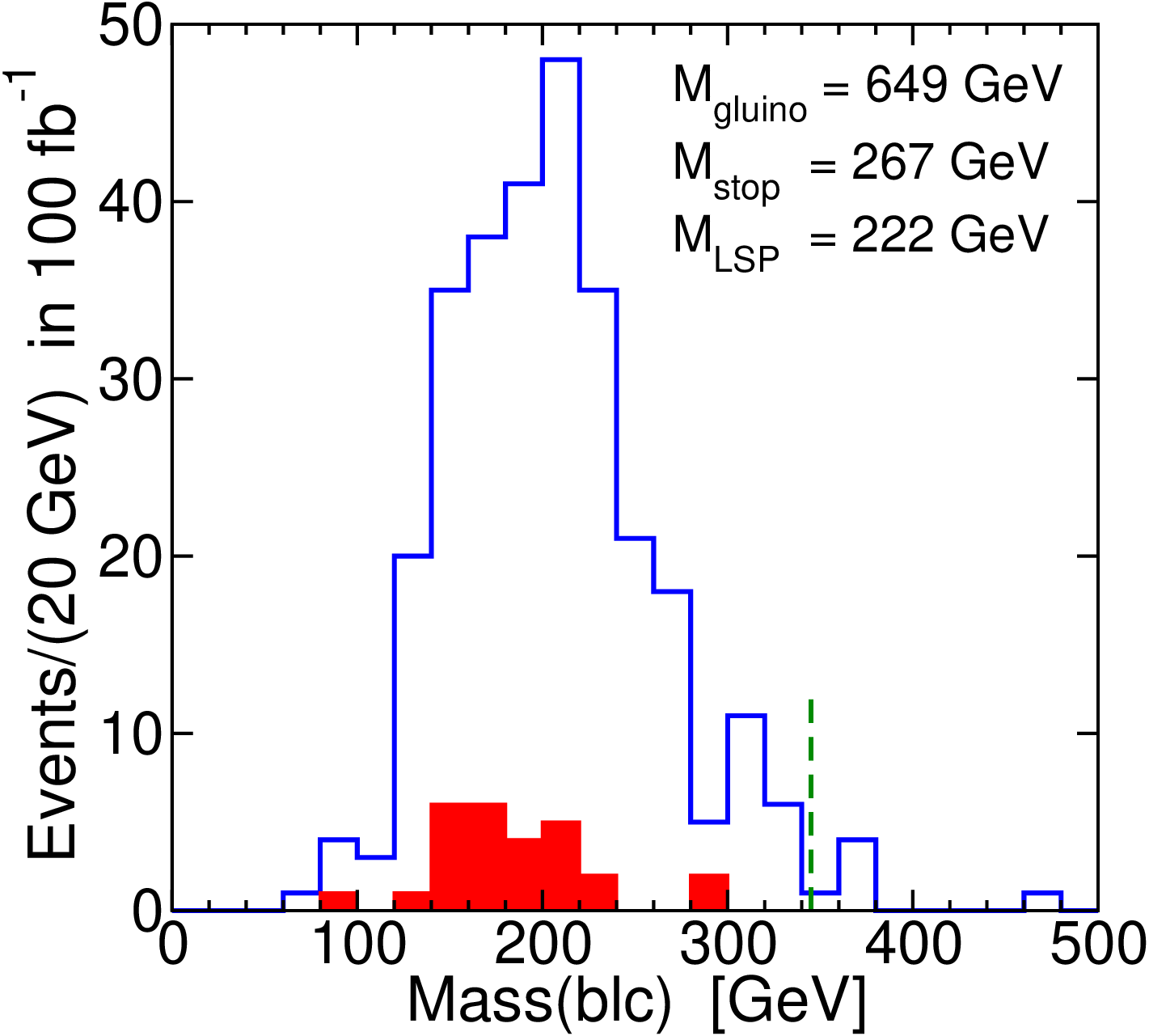}
\caption{\label{fig:massblc}
The distributions of 
$m(b\ell c)$ for six points on the model line 
described in section \ref{sec:modelline} with 
$M_1 =$ 425, 450, 475, 500, 525, and 550 GeV. 
The nominal endpoints, indicated by the dashed vertical lines, are given
by eq.~(\ref{eq:mbcmax}) as, respectively, 
$m(b\ell c)_{\rm max} =$  273, 339, 350, 354, 350, and 345 GeV.
The solid histograms show the portion of the signal for which
the putative $c$ jet has a heavy flavor tag or a soft muon tag.}
\end{figure}
\begin{figure}[!tp]
\includegraphics[width=7.93cm,angle=0]{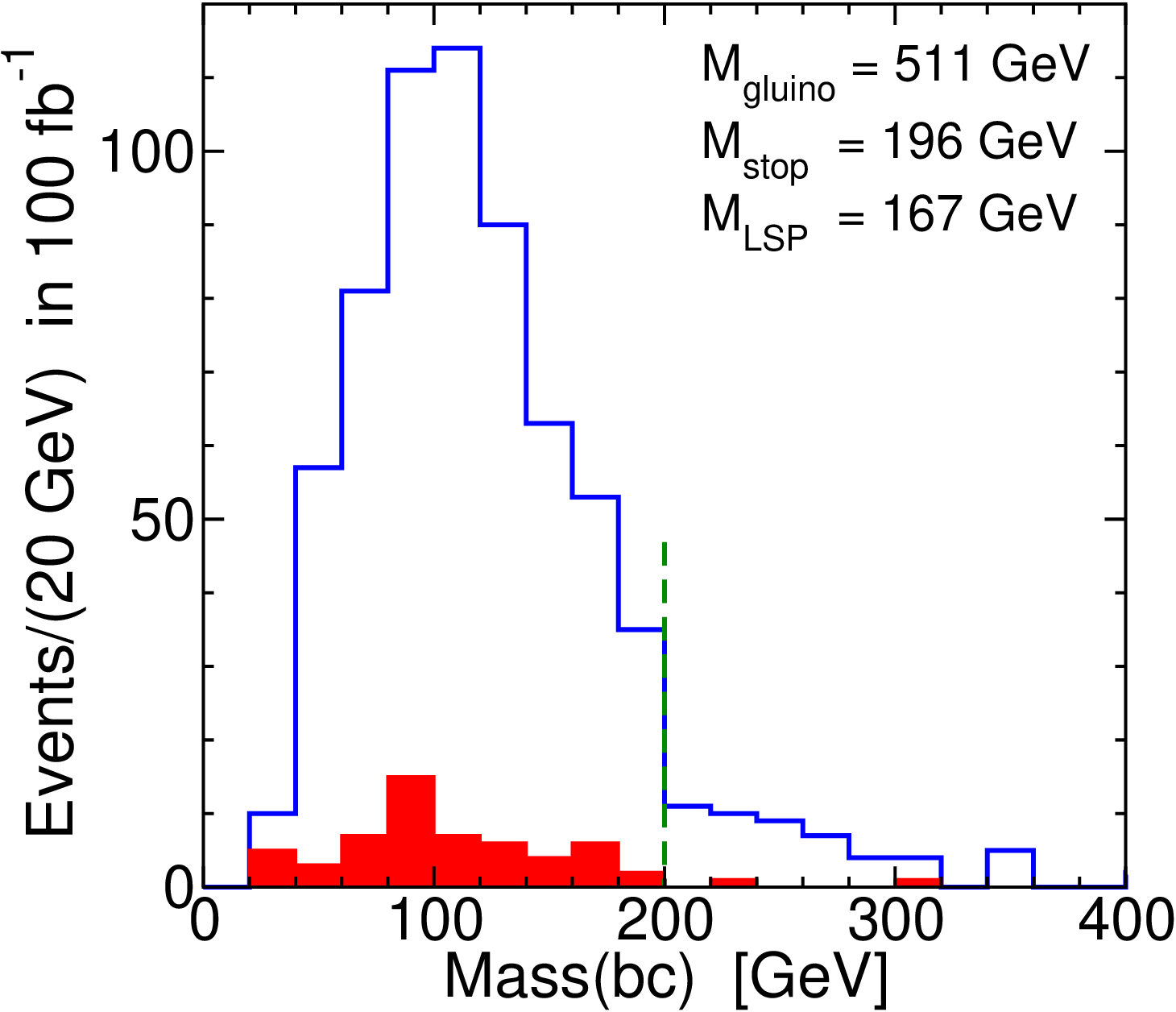}~~~
\includegraphics[width=7.93cm,angle=0]{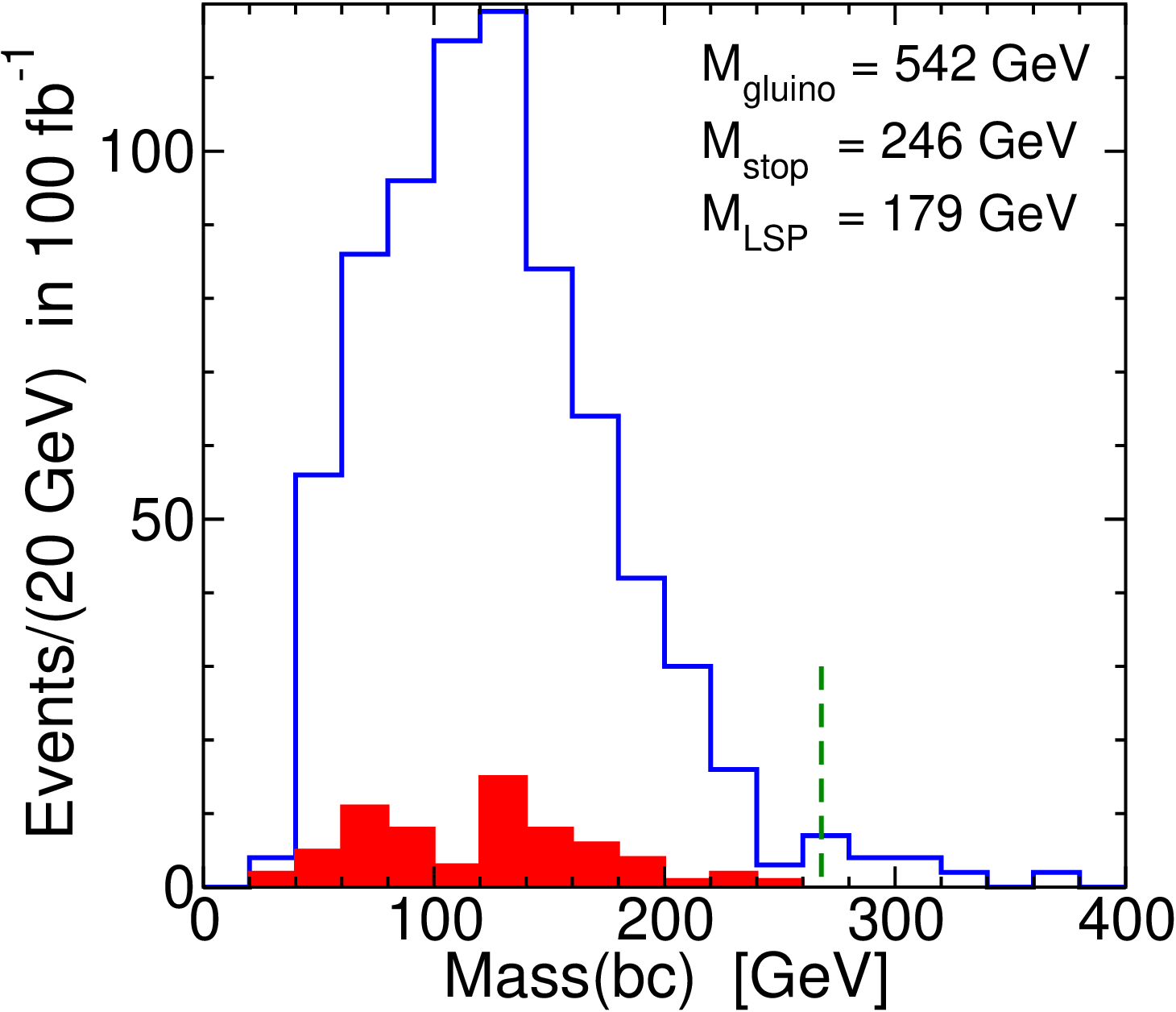}
\\
\includegraphics[width=7.93cm,angle=0]{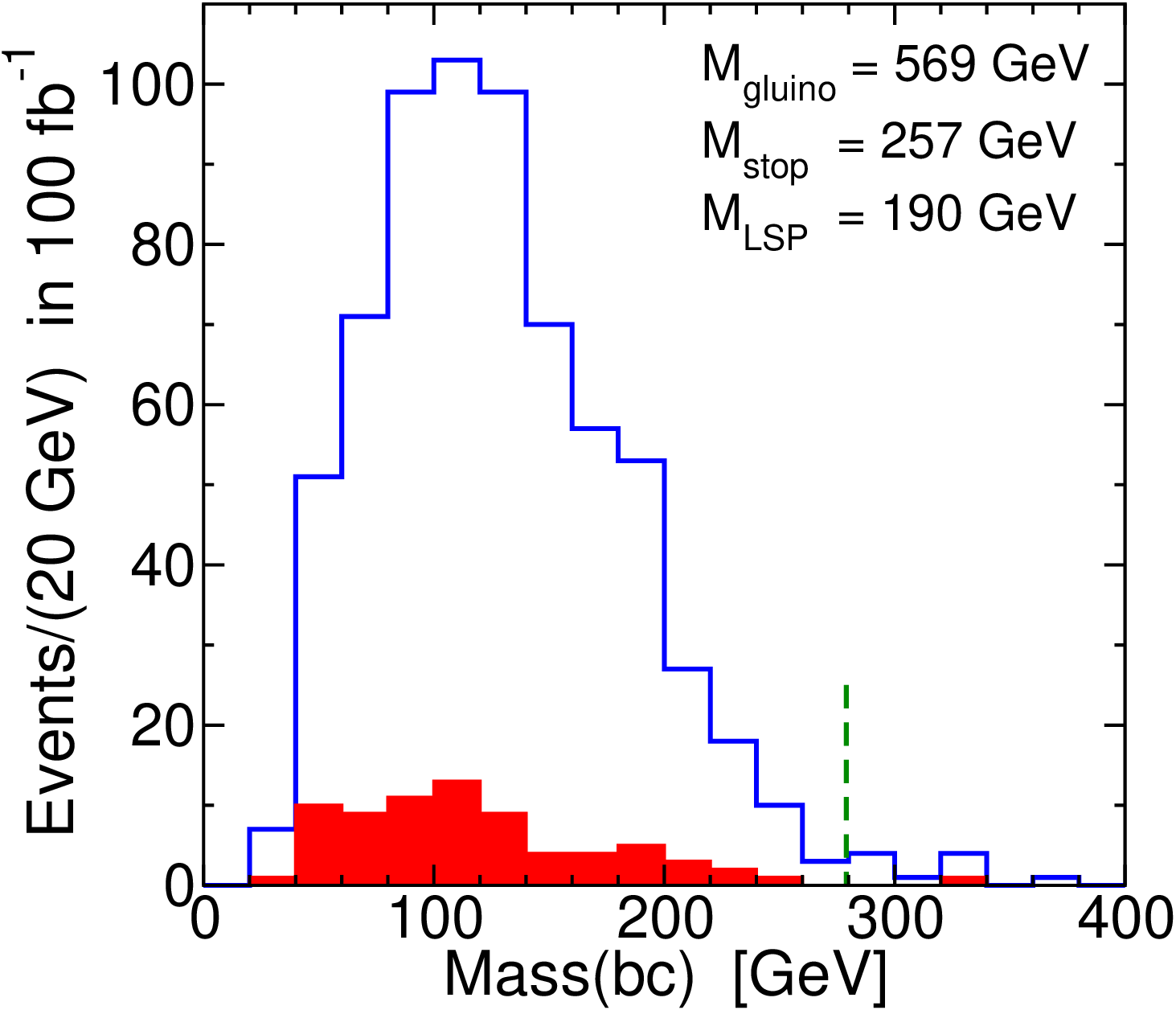}~~~
\includegraphics[width=7.93cm,angle=0]{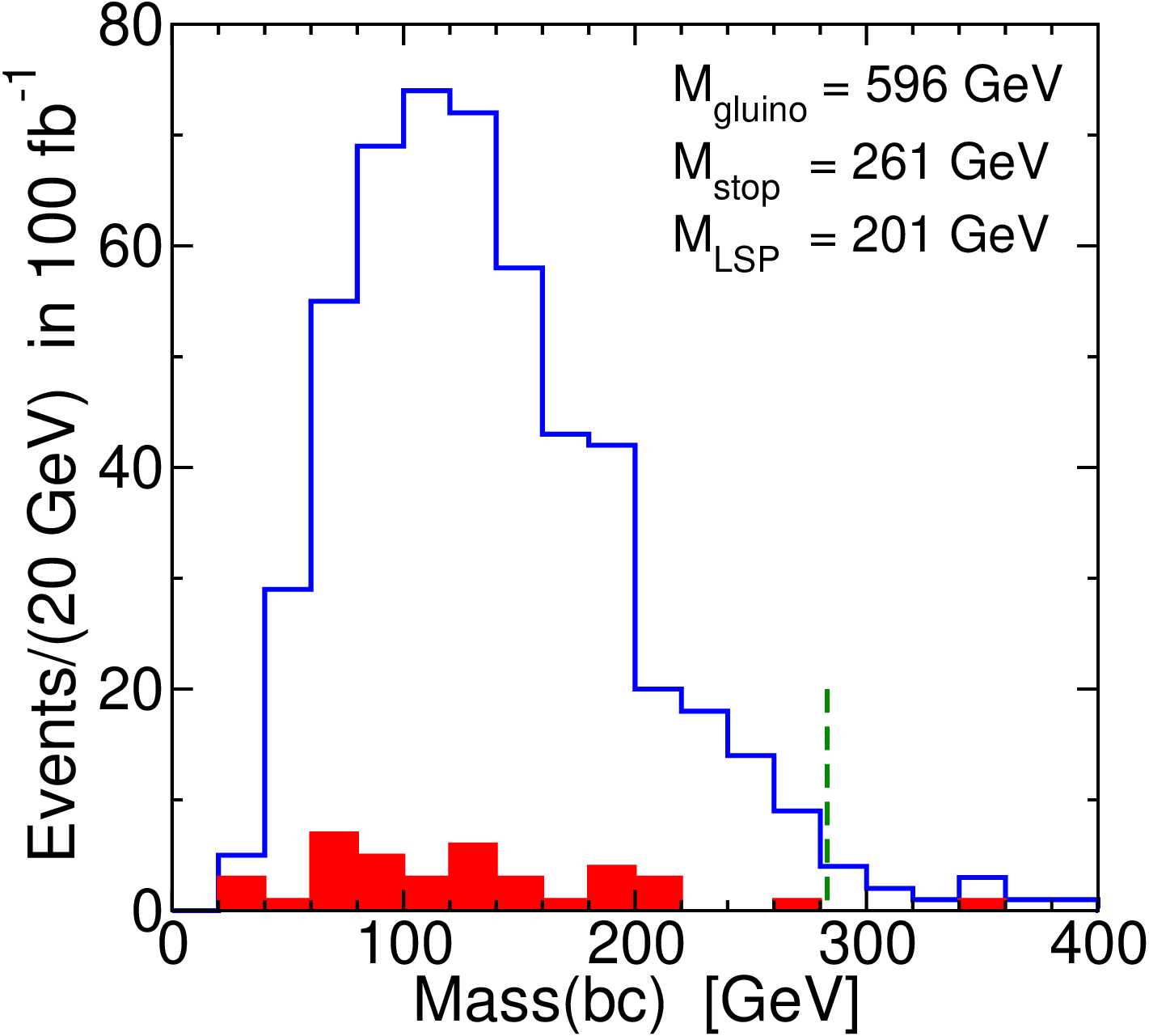}
\\
\includegraphics[width=7.93cm,angle=0]{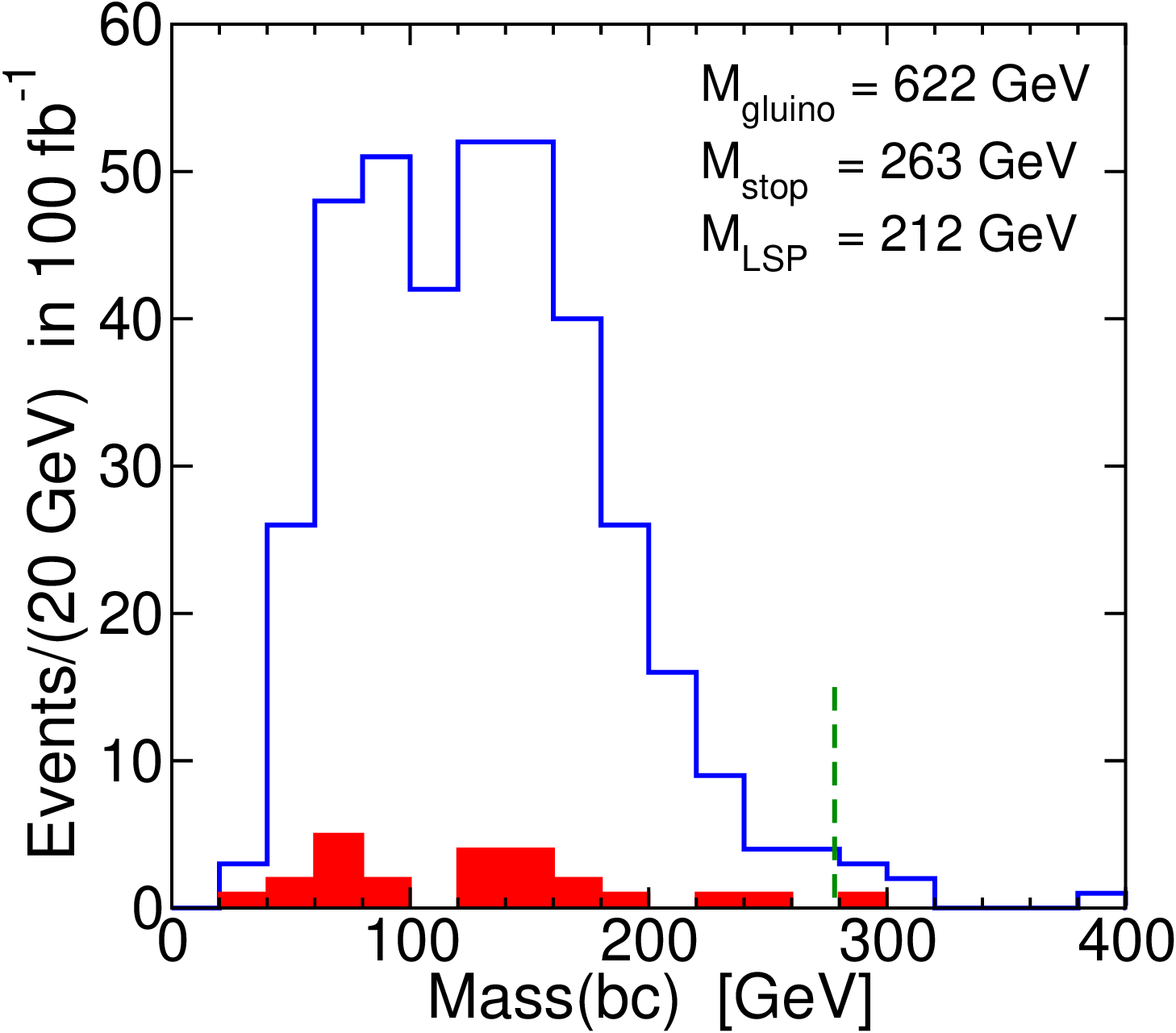}~~~
\includegraphics[width=7.93cm,angle=0]{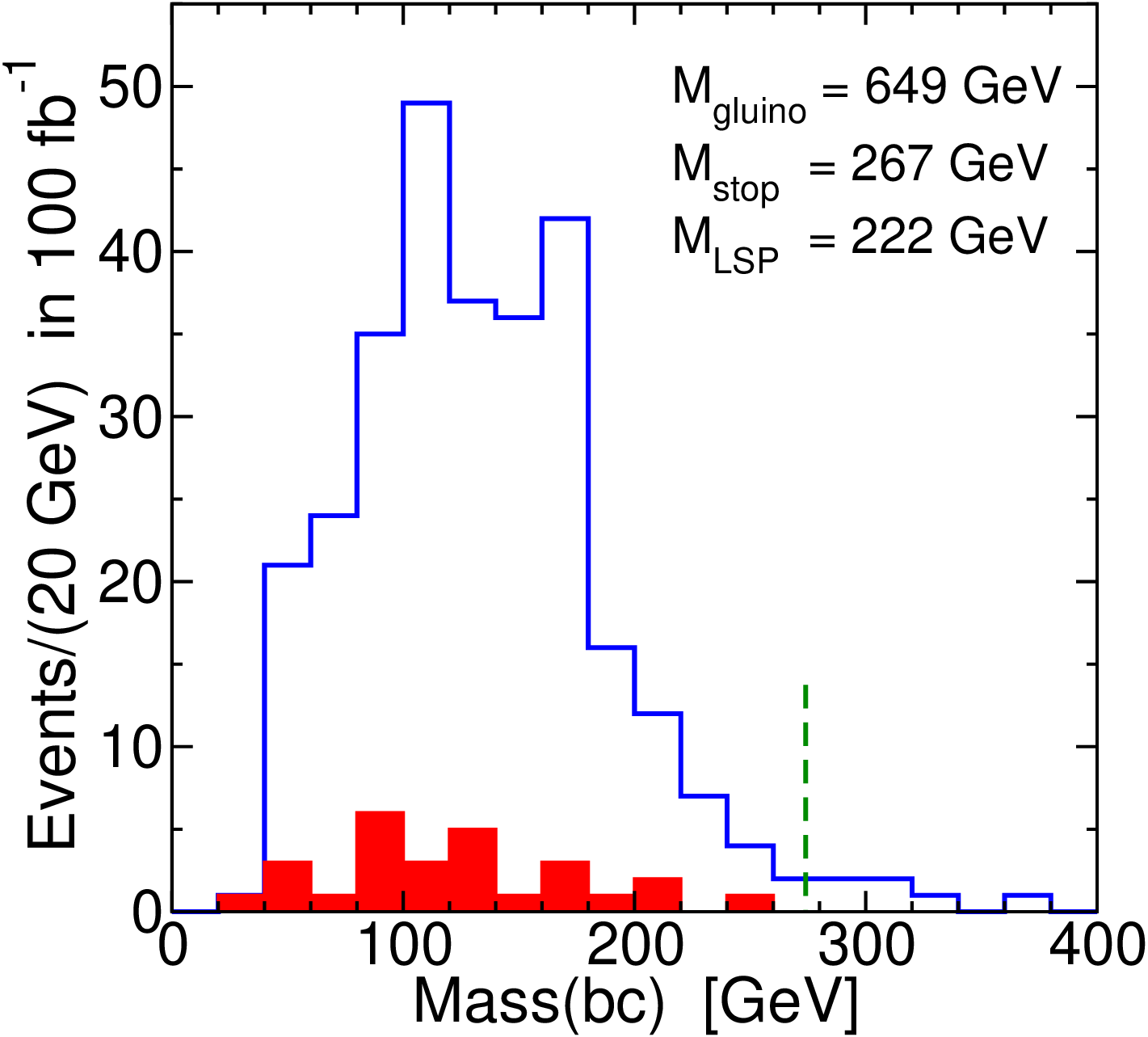}
\caption{\label{fig:massbc}
The distributions of 
$m(bc)$ for six points on the model line 
described in section \ref{sec:modelline} with 
$M_1 =$ 425, 450, 475, 500, 525, and 550 GeV. 
The nominal endpoints, indicated by the dashed vertical lines, are given
by eq.~(\ref{eq:mbcmax}) as, respectively, 
$m(bc)_{\rm max} =$ 200, 268, 279, 283, 278, and 274 GeV.
The solid histograms show the portion of the signal for which
the putative $c$ jet has a heavy flavor tag or a soft muon tag.}
\end{figure}
Results for the $m(b\ell c)$ and $m(bc)$ distributions selected in this
way are shown in figures \ref{fig:massblc} and \ref{fig:massbc} for
several representative models. These distributions are seen to be roughly
consistent with endpoints at the nominal positions, but wrong assignments
and jet energy smearing leads to some events in a high-mass tail in each
case. This can be seen to be particularly troublesome for the lowest-mass
$M_1 = 425$ GeV ($m_{\tilde g} = 511$) model point, where the small mass
difference $m_{\tilde t_1} - m_{\tilde N_1} = 29$ GeV means that the true
charm jets often fail the $p_T>35$ GeV cut.\footnote{I have checked that
lowering this cut does not help significantly, because doing so also
allows more interloper jets.} This exemplifies a more general difficulty.
If the thermal relic abundance of neutralinos does not account for all of
the dark matter, then the stop-LSP mass difference will be smaller than
indicated in figure \ref{fig:massdiff}, for any given gluino masses. This
can always lead to the problem of the charm jets having too small $p_T$
and being replaced in the analysis by interlopers, leading to a distorted
distribution and a tail above the true mass endpoint. To counteract this
problem, one could use an independent check on the identity of the charm
jet. In figures \ref{fig:massblc} and \ref{fig:massbc}, the solid
histograms show the portion of the signal for which the putative charm jet
has a PGS heavy-flavor tag or contains a non-isolated muon (similar to the
``soft muon" tag used in Fermilab Tevatron analyses). This information
will clearly be more useful if the efficiency and purity of ``charm
tagging" can be improved. Although I will not attempt it here in the
absence of a fully realistic detector simulation, one can imagine that a
likelihood fit taking into account these effects could give measurements
(or at least constraints) on these endpoints. However, figures
\ref{fig:massblc} and \ref{fig:massbc} show that precision may be 
difficult to achieve without either more data than 
100 fb$^{-1}$ or a better handle on charm jets. 

\begin{figure}[!tp]
\begin{minipage}[]{0.37\linewidth}
\caption{\label{fig:endpoints}
The parton-level prediction for the 
endpoints of the $m(b\ell c)$ and $m(bc)$ distributions,
for points along the model line described in section \ref{sec:modelline}, 
as a function
of the gluino mass.}
\end{minipage}
\begin{minipage}[]{0.62\linewidth}
\begin{flushright}
\includegraphics[width=9.0cm,angle=0]{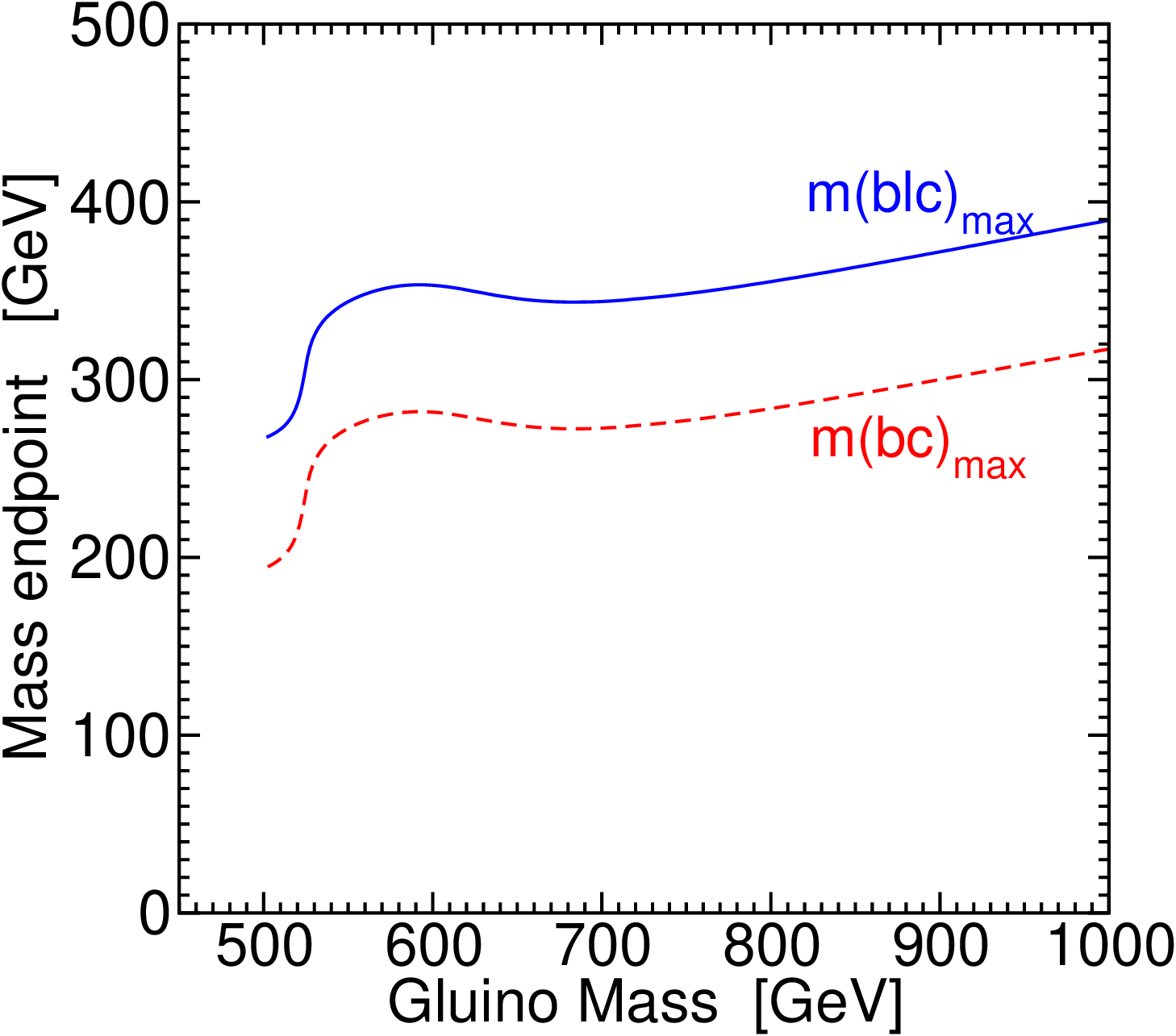}
\end{flushright}
\end{minipage}
\end{figure}

Unfortunately, even with such a measurement, most of these models are
quite indistinguishable from each other using the endpoints or shapes of
the distributions alone.  This is because the endpoints are nearly
independent of the mass scale defining the point along the model line
studied here, as illustrated in figure \ref{fig:endpoints}.%
In fact, for the entire range 545 GeV $<m_{\tilde g} < $ 830 GeV, 
$m(b\ell c)_{\rm max}$ is within 10 GeV of 350 GeV for this model line.  
(For lower values of the gluino mass, the endpoint is lower, but its
determination becomes much more problematic due to wrong jet assignments
due to the smaller stop-LSP mass difference, as we have just seen.) It
might at first seem surprising that the position of the endpoints does not
scale with the gluino mass. The reason is that the scaling is counteracted
by the factor of $(1 - m^2_{\tilde N_1}/m_{\tilde t_1}^2)$ in the formulas
eqs.~(\ref{eq:mlcmax})-(\ref{eq:mblcmax}), which decreases as one moves to
higher masses along the model line, because of the constraint on the
stop-LSP mass difference coming from the dark matter abundance
observation. 

\begin{figure}[!tp]
\begin{minipage}[]{0.37\linewidth}
\caption{\label{fig:gluinostop}
The relation between the gluino ($\tilde g$) and 
the lighter stop ($\tilde t_1$) masses following
from eq.~(\ref{eq:mblcmax}),
for the case that the high endpoint of the $m(b\ell c)$
distribution is taken to be 350 GeV. The different lines correspond to
$m_{\tilde t_1} - m_{\tilde N_1} = 30$, 45, 60, and 85 GeV.
The last is the maximum mass difference that allows 
$\tilde t_1 \rightarrow c \tilde N_1$ to dominate, as required by the
signal.}
\end{minipage}
\begin{minipage}[]{0.62\linewidth}
\begin{flushright}
\includegraphics[width=9.0cm,angle=0]{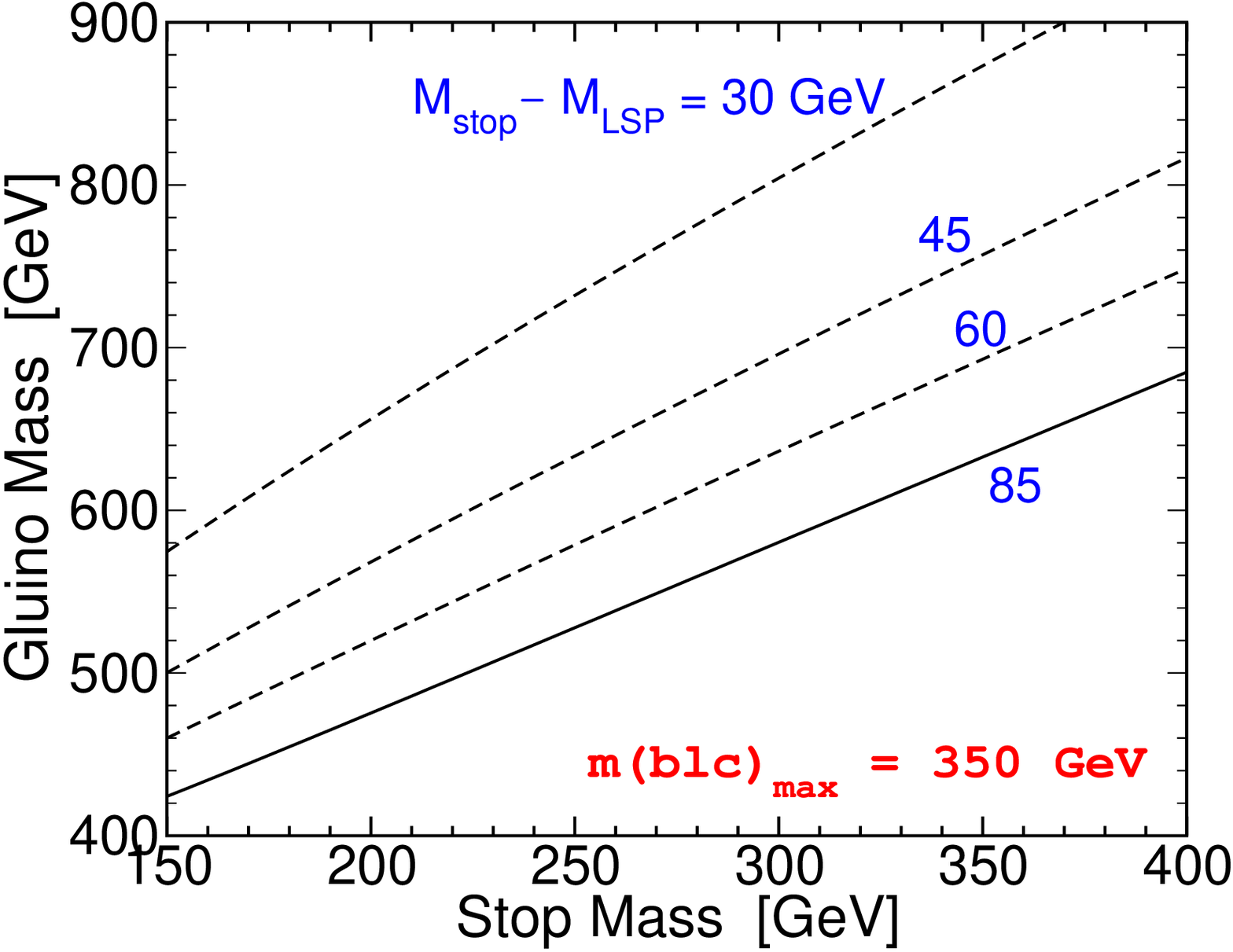}
\end{flushright}
\end{minipage}
\end{figure}

Nevertheless, a successful determination of the endpoints
will still be useful when combined with the
information that the gluino decay signal is kinematically allowed at all.
This is illustrated in figure \ref{fig:gluinostop}, which assumes that the
$m(b\ell c)$ endpoint is found to be 350 GeV (or equivalently that the 
$m(bc)$ endpoint is found at 279 GeV).%
The allowed line in the gluino mass vs. stop mass plane is shown, for
various assumptions about the stop-LSP mass difference. For the signal to
occur at all, one must have $m_{\tilde t_1} - m_{\tilde N_1} < 85$ GeV,
otherwise the decay $\tilde t_1 \rightarrow c \tilde N_1$ would lose to
the flavor-preserving three-body decay $\tilde t_1 \rightarrow b W \tilde
N_1$.  (As noted at the end of section \ref{sec:sstops}, this can be ruled
out by counting the number of additional heavy-flavor tags in the events
that pass the signal selections.) This means that for a given stop mass,
the gluino mass must be above the solid line. The dashed lines show the
gluino-stop mass relation for smaller values of the $m_{\tilde t_1} -
m_{\tilde N_1}$ mass difference.  [If the $m(b\ell c)$ endpoint is only
constrained to be $\geq 350$ GeV, then the allowed regions are above the
indicated lines.] Now, combining this information with an estimate or
upper bound on the gluino mass from the production cross-section or from
the observables of the type $M_{A,B,C,D}$ described above would allow a
determination of ranges in which the gluino, stop, and LSP masses must be. 

\section{Outlook}
\label{sec:outlook}
\setcounter{equation}{0}
\setcounter{footnote}{1}

Compressed supersymmetry with top-squark mediation of neutralino
annihilation in the early universe presents both challenges and
opportunities for the LHC. Although early discovery should not be a
problem because of the low mass scale, the sleptons and the charginos and
neutralinos (other than the LSP) may very nearly decouple.  In the model
line studied here, for example, it is very difficult and perhaps
impossible for the LHC to be able to say anything about them. With
sufficient integrated luminosity, one may be able to discover stoponium
through its diphoton decays \cite{Drees:1993yr,Martin:2008sv}, giving a
uniquely precise measurement of the top-squark mass. This would provide an
important absolute reference point for determination of the other
superpartner masses. However, it requires that the top-squark mass is not
too large. 

Other than stoponium, the most distinctive signature may be the same-sign
leptonic top-quark decays that come from gluinos (or squarks decaying to
gluinos).  In this paper, I have studied the prospects for learning about
the gluino, top-squark and LSP masses from these events. 

First, the scenario is distinguishable from similar ones with a larger
stop-LSP mass difference by using the frequency of additional heavy-flavor
tags in the $\ell^\pm\ell^\pm bb jj + \missET$ events after cuts. 

Observables obtained by summing over scalar $p_T$'s and $\missET$ within
this relatively clean sample will provide estimates of the effective
superpartner mass scale, which is always strongly correlated with the
gluino mass. This can be compared with the estimate obtained from $H_T$
and $M_{\rm eff}$ distributions in the larger inclusive jets+$\missET$
sample. 

The determination of invariant mass endpoints is somewhat more
problematic, due to the pernicious effects of interloping jets (both from
squark decays and from the underlying event) being confused with the charm
jet in the analysis. In the actual LHC analysis, this can probably be
enhanced by using heavy-flavor likelihoods on an event-by-event basis to
help choose the correct charm jets. These endpoints do not provide
unambiguous information about the superpartner masses, even within the
confines of the single model line studied here. However, when combined
with the information that the decay $\tilde t_1 \rightarrow c \tilde N_1$
dominates, this information can also be useful to constrain the model.
Clearly, heavy flavor tagging will be crucial in this effort. Also, if one
is willing to assume that the thermal relic abundance of dark matter is
due entirely to neutralino LSPs without fine-tuning, then the resulting
stop-LSP mass difference should be large enough to more sharply define
the endpoints. Conversely, a confirmation of this scenario would help to
establish the supersymmetric interpretation of the dark matter. 

\bigskip \noindent 
{\it Acknowledgments:} 
I am indebted to John Conway, Dave Hedin, Michel Herquet, Gudrun Hiller, Tilman Plehn,
and Tim Stelzer for helpful communications. 
This work was supported in part by the National Science Foundation grant 
number PHY-0456635.
This research was supported in part by the National Science Foundation 
under Grant No. PHY05-51164.


\end{document}